\documentstyle[preprint,eqsecnum,aps,epsf,equation]{revtex}
\newif\iftightenlines\tightenlinesfalse
\tightenlines\tightenlinestrue
\def\eslt{E\llap/_T}
\def\etmiss{E\llap/_T}
\def\to{\rightarrow}
\def\te{\tilde e}

\def\tb{\tilde b}
\def\tf{\tilde f}
\def\td{\tilde d}
\def\tst{\tilde t}
\def\tt{\tilde t}
\def\ttau{\tilde \tau}

\def\tg{\tilde g}
\def\tnu{\tilde\nu}

\def\tq{\tilde q}
\def\tw{\widetilde W}
\def\tz{\widetilde Z}

\def\sgn{\mathop{\rm sgn}}
\def\mhf{m_{1/2}}

\def\ETC{E_T^c}
\newcommand{\be}{\begin{equation}}
\newcommand{\ben}{\begin{subequations}}
\newcommand{\een}{\end{subequations}}
\newcommand{\beq}{\begin{eqalignno}}
\newcommand{\eeq}{\end{eqalignno}}
\newcommand{\ee}{\end{equation}}
\newcommand{\wt}{\widetilde}

\newcommand{\tanb}{\mbox{$\tan \! \beta$}}

\newcommand{\cosb}{\mbox{$\cos \! \beta$}}
\newcommand{\sinb}{\mbox{$\sin \! \beta$}}

\begin{document}
%
\preprint{\vbox{\baselineskip=14pt%
   \rightline{FSU-HEP-980204}\break 
   \rightline{BNL-HET-98/9}\break 
   \rightline {APCTP 98-02}
   \rightline{UH-511-894-98}
}}
\title{SUPERSYMMETRY REACH OF TEVATRON UPGRADES:\\
THE LARGE $\tan\beta$ CASE}
\author{Howard Baer$^1$, Chih-Hao Chen$^2$, Manuel Drees$^3$, 
Frank Paige$^4$\\
 and Xerxes Tata$^{5}$}
\address{
$^1$Department of Physics,
Florida State University,
Tallahassee, FL 32306 USA
}
\address{
$^2$Department of Physics,
University of California,
Davis, CA USA
}
\address{
$^3$APCTP, 207-43 Cheongryangri-dong, Seoul 130-012, Korea
}
\address{
$^4$Brookhaven National Laboratory, 
Upton, NY 11973 USA
}
\address{
$^5$Department of Physics and Astronomy,
University of Hawaii,
Honolulu, HI 96822 USA
}
\date{\today}
\maketitle
\begin{abstract}

The Yukawa couplings of the tau lepton and the bottom quark become 
comparable to, or even exceed, electroweak gauge couplings for large 
values of the SUSY parameter $\tan\beta$. 
As a result, the
lightest tau slepton $\ttau_1$ and bottom squark $\tb_1$ can be significantly 
lighter than corresponding sleptons and squarks of the first two
generations. Gluino, chargino and neutralino decays to third generation
particles are significantly enhanced when $\tan\beta$ is large.
This affects projections for collider experiment
reach for supersymmetric particles.
In this paper, we evaluate the reach of the Fermilab Tevatron $p\bar p$
collider for supersymmetric signals in the framework of the mSUGRA model.
We find that the reach via 
signatures with multiple isolated leptons ($e$ and $\mu$) is
considerably reduced.
For very large $\tan\beta$, the greatest reach is 
attained in the multi-jet$+\eslt$ signature.
Some significant extra regions may be probed by requiring the presence of 
an identified $b$-jet in jets$+\eslt$ events, or by requiring 
one of the identified leptons in clean trilepton events to actually be 
a hadronic 1 or 3 charged prong tau. In an appendix, we present 
formulae for chargino, neutralino and gluino three body 
decays which are valid at large $\tan\beta$.

\end{abstract}

\medskip



\section{Introduction and Motivation}

The minimal supergravity model (mSUGRA)~\cite{sugra} is commonly regarded as 
the paradigm framework for phenomenological analyses of weak scale
supersymmetry. The visible sector is taken to consist of the particles
of the Minimal Supersymmetric Standard Model\cite{mssm}
(MSSM). 
One posits, in addition, the existence of
``hidden sector'' field(s), which couple to ordinary matter fields
and their superpartners only
via gravity. The conservation of $R$-parity is assumed.
Supersymmetry is broken 
in a hidden sector of the theory;
supersymmetry breaking is then communicated to the visible sector via
gravitational interactions. The technical 
assumption of minimality implies
that kinetic terms for matter fields take the canonical form; 
this assumption, which is equivalent to assuming an approximate global $U(n)$
symmetry between $n$ chiral multiplets, leads to a common mass squared 
$m_0^2$ for all 
scalar fields, and a common trilinear term $A_0$ for all $A$ parameters.
These parameters, which determine the sparticle-particle mass splitting
in the observable sector are taken to be comparable to the weak scale,
$M_{weak}$ .
In addition, motivated by the apparently successful gauge coupling 
unification in the MSSM,
one usually adopts a common value $m_{1/2}$ for all gaugino masses at the
scale $M_{GUT}\simeq 2\times 10^{16}$ GeV. For simplicity, it is 
commonly assumed that in fact the scalar masses and trilinear terms unify at
$M_{GUT}$ as well. The resulting effective theory, valid at energy scales
$E<M_{GUT}$, is then just the MSSM with the usual soft SUSY breaking terms,
which in this case are unified at $M_{GUT}$. 
The soft SUSY breaking scalar and gaugino mases, the
trilinear $A$ terms and in addition a bilinear soft term $B$, the gauge and
Yukawa couplings and the supersymmetric $\mu$ term are all then evolved
from $M_{GUT}$ to some scale $M\simeq M_{weak}$ using renormalization 
group equations (RGE's). 
The large top quark Yukawa coupling causes the squared mass of one of the 
Higgs fields to be driven negative, resulting in the 
breakdown of electroweak symmetry; this determines the value 
of $\mu^2$. Finally, it is customary to trade the parameter $B$ for
$\tan\beta$, the ratio of Higgs field vacuum expectation values.
The resulting weak scale spectrum of superpartners and their couplings can
thus be derived in terms of four continuous plus one discrete
parameters
\begin{equation}
m_0,\ m_{1/2},\ A_0,\ \tan\beta\ {\rm and} \sgn(\mu),
\end{equation}
in addition to the usual parameters of the standard model.

The consequences of the mSUGRA model have been investigated for collider
experiments at the CERN LEP2 $e^+e^-$ collider\cite{lep2}, the Fermilab
Tevatron $p\bar p$ collider\cite{bcpt,tevatron}, the CERN LHC $pp$
collider\cite{lhc} and a possible next linear $e^+e^-$ collider (NLC)
operating at $\sqrt{s}\simeq 500$ GeV\cite{nlc,NOJ}. In all
but the last of these studies (where the effect of the tau Yukawa
coupling on aspects of the phenomenology of the stau sector is carefully
examined), small to moderate values of the parameter $\tan\beta\sim
2-10$ have been adopted.  This was due in part to the fact that event
generators such as ISAJET\cite{isajet} had not been constructed to
provide reliable calculations for large $\tan\beta$. In particular,
effects of tau and bottom Yukawa couplings,
\begin{equation}
 f_b={g m_b\over\sqrt{2}M_W\cos\beta},
\ f_{\tau}={g m_{\tau}\over\sqrt{2}M_W\cos\beta}
\end{equation}
which become comparable to the electroweak gauge couplings and even to
the top Yukawa coupling $f_t=g m_t/(\sqrt{2}M_W\sin\beta)$ if
$\tan\beta$ is large, had not been completely included.  The correct
inclusion of these couplings has a significant impact~\cite{drees,prl}
on the search for supersymmetry at colliders.

In the mSUGRA model, the parameter $\tan\beta$ can be as large as
$\tan\beta\sim {m_t/m_b}$, where the quark masses are evaluated at
a scale $\sim M_{weak}$; since the running $m_b$ is considerably smaller
than 5~GeV, $\tan\beta$ values up to 45-50 are possible.  Such large
$\tan\beta$ values are indeed preferred in some $SO(10)$ GUT models with
Yukawa coupling unification. In practice, one finds that if $\tan\beta$
is chosen to be too large, $f_b$ diverges before $M_{GUT}$. A slightly
stronger upper limit on $\tan\beta$ is obtained from the requirement
that $m_A^2$, the mass of the pseudo-scalar Higgs boson, should be
positive. The precise value of the upper bound on $\tan\beta$ 
depends somewhat on the other mSUGRA parameters.
%

In a recent Letter\cite{prl},
we reported on an upgrade of the event generator ISAJET 
that correctly incorporated the effects of $\tau$ and $b$ Yukawa interactions
so that it
would provide reliable predictions for 
supersymmetry with large $\tan\beta$.
Novel phenomenological implications special to 
large values of 
$\tan\beta$ were pointed out: in particular, it was noted that while
Tevatron signals in multilepton ($e$ and $\mu$) channels were
greatly reduced, there could be new signals involving $b$-jets and
$\tau$-leptons via which
to search for SUSY.
In this paper, we focus our attention on the search for
supersymmetry at the Main Injector (MI) upgrade of 
the Fermilab Tevatron $p\bar p$
collider,
($\sqrt{s}=2$ TeV, integrated luminosity $\int{\cal L}dt=2~$fb$^{-1}$)
and the proposed TeV33 upgrade 
($\sqrt{s}=2$ TeV, integrated luminosity $\int{\cal L}dt=25~$fb$^{-1}$)
for the case where $\tan\beta$ is large.

\subsection{Sparticles masses at large $\tan\beta$}

Large $b$ and $\tau$ Yukawa couplings
significantly  alter the mass spectra of the sparticles and Higgs bosons as
shown in Fig.~1. Here we plot various sparticle and Higgs boson
masses versus $\tan\beta$ for  mSUGRA parameters
$m_{1/2}=150$ GeV, $A_0=0$ and {\it a}) $m_0=150$ GeV and {\it b})
$m_0=500$ GeV, for both signs of $\mu$. We fix the pole mass $m_t = 170$~GeV.

The $b$ and $\tau$ Yukawa couplings contribute negatively to the
renormalization group running of the sbottom and stau soft masses, driving 
them to lower values than soft masses for the corresponding 
first and second generation squarks and sleptons. In addition, 
the off-diagonal terms in the sbottom and stau mass-squared matrices
$m_b(-A_b+\mu\tan\beta$) and $m_{\tau}(-A_{\tau}+\mu\tan\beta$)
can result in significant mixing between 
left and right sbottom and stau gauge eigenstates, 
and a possible further decrease in
the physical masses for the lighter of the two sbottom (and stau)
mass eigenstates $m_{\tb_1}$ and $m_{\ttau_1}$. 
If $\tan\beta$ is small, $\ttau_1 \simeq \ttau_R$, while (because of top
quark Yukawa interactions) $\tb_1 \simeq \tb_L$.
The impact of bottom and tau Yukawa interactions can be seen
in Fig.~1: $m_{\ttau_1}\simeq m_{\te_R}$ at low $\tan\beta$, and
as $\tan\beta$ increases, $m_{\ttau_1}$ decreases, while $m_{\te_R}$
remains constant. Likewise, $m_{\tb_1}$ decreases with increasing
$\tan\beta$, while $m_{\td_L}$ remains constant. In the case of 
frame {\it a}), ultimately $m_{\ttau_1}$ drops below $m_{\tw_1}$ and $m_{\tz_2}$
so that the two body decays $\tw_1\to \ttau_1\nu_\tau$ and
$\tz_2\to\ttau_1\tau$ become allowed, and dominate the
branching fractions.

It is well known that at low to moderate values of $\tan\beta$, 
the large top Yukawa coupling drives the Higgs
mass $m_{H_2}^2$ to negative values, resulting in a breakdown of electroweak
symmetry. At large $\tan\beta$, the large $b$ and $\tau$ Yukawa couplings
drive the other soft Higgs mass-squared $m_{H_1}^2$ to small or negative values
as well. This results overall in a {\it decrease} 
in mass for the pseudo-scalar
Higgs $m_A$ relative to its value at small $\tan\beta$. Since the values of the
heavy scalar and charged Higgs boson masses are related to $m_A$, 
they decrease as well. This effect is also illustrated in Fig.~1, where
the mass $m_A$ decreases dramatically with increasing $\tan\beta$.
The curves are terminated at the value of $\tan\beta$ beyond which
$m_A^2
< 0$, and the correct pattern of electroweak symmetry breaking is not
obtained as already mentioned.
We found that the pseudoscalar mass $m_A$, obtained using the 1-loop
effective potential, is unstable by up to factors of two
against scale variations for relatively low values of scale choice
$Q\sim M_Z$.
This instability would be presumably corrected by inclusion of
2-loop corrections.
We find the choice of scale
$Q\sim\sqrt{m_{\tst_L}m_{\tst_R}}$ to empirically yield stable predictions of
Higgs boson masses in the RG improved 1-loop effective potential
(where we include contributions from all third generation
particles and sparticles).
This scale choice effectively includes some
important two loop effects, and yields predictions for light scalar Higgs boson
masses $m_h$ in close accord with the results of Ref. \cite{carena}.

\subsection{Sparticle decays at large $\tan\beta$}

For large values of $\tan\beta$, $b$ and $\tau$ Yukawa couplings become
comparable in strength to the usual gauge interactions, so that Yukawa
interaction
contributions to sparticle decay rates are non-negligible and can even
dominate. This could manifest itself as lepton non-universality in SUSY
events. Also, because of the reduction of masses referred to
above, chargino and neutralino decays to stau, sbottom
and various Higgs bosons
may be allowed, even if the corresponding decays would be
kinematically forbidden for small $\tan\beta$ values.
The reduced stau, sbottom, and Higgs masses can also
increase sparticle branching ratios to third generation particles
via virtual effects. These enhanced decays to third generation
particles can radically alter
the expected SUSY signatures at colliders.

We have
re-calculated the branching fractions for the $\tg$, $\tb_i$, $\tst_i$,
$\ttau_i$, $\tnu_{\tau}$, $\tw_i$, $\tz_i$, $h$, $H$, $A$ and $H^\pm$
particles and sparticles including sbottom and stau mixing as well as
effects of $b$ and $\tau$ Yukawa interactions.
For Higgs boson decays, we use the formulae in Ref. \cite{bisset}.
We have recalculated the decay widths for
$\tg\to tb\tw_i$ and $\tg\to b\bar{b}\tz_i$. These have been 
calculated previously
by Bartl {\it et al.}\cite{bartl}; our results agree with theirs if we
use pole fermion masses to calculate
the Yukawa couplings. In ISAJET, we use the
running Yukawa couplings evaluated at the scale $Q=m_{\tg}$ ($m_t$) to compute
decay rates for the gluino ($\tw_i$,$\tz_i$). This seems a more
appropriate choice, and it significantly alters
the decay widths when effects of $f_b$ are important.
The $\tz_i\to \tau\bar{\tau}\tz_j$ and $\tz_i\to b\bar{b}\tz_j$
decays take place via eight diagrams ($\tf_{1,2}$,
$\bar{\tf}_{1,2}$, $Z$, $h$, $H$ and $A$ exchanges). In our
calculation of $\tg$ and $\tz_i$ decays, 
we have neglected $b$ and $\tau$ masses except
in the Yukawa couplings and in the phase space integration.
We have also computed
the widths for decays $\tw_i\to\tz_j \tau\nu$ which are mediated by
$W$, $\ttau_{1,2}$, $\tnu_{\tau}$ and $H^{\pm}$ exchanges; in these cases,
we retain $m_{\tau}$ effects only in the Yukawa couplings. Formulae for
these three-body decays are presented in the Appendix.

To illustrate the importance of the Yukawa coupling effects,
we show selected branching ratios of
$\tw_1$ and $\tz_2$ in Fig.~2.
In all frames we take $\mu >0$.
Frames {\it a})
and {\it b}) are for the mSUGRA case
($m_0,\ m_{1/2},\ A_0 )=(150,150,0)$ GeV; frames {\it c}) and {\it d})
show the same branching fractions, but take $m_0=500$ GeV instead.
In frame {\it a}), for low $\tan\beta$
we see that the $\tw_1\to e\nu\tz_1$ and $\tw_1\to\tau\nu\tz_1$
branching ratios are very close in magnitude, reflecting the smallness
of $f_{\tau}$. For $\tan\beta \agt 10$, these
branchings begin to diverge, with the branching to $\tau$'s
becoming increasingly
dominant. For $\tan\beta >40$, the two body mode $\tw_1\to \ttau_1\nu$
opens up and quickly dominates. Since this decay
is followed by $\ttau_1\to \tau\tz_1$, the end products of chargino
decays here are almost exclusively
tau leptons plus missing energy.

In frame {\it b}), we see at low $\tan\beta$ the $\tz_2\to
e\bar{e}\tz_1$ and $\tz_2\to \tau\bar{\tau}\tz_1$ branchings are large
($\sim 10\%$) and equal, again because of the smallness of the Yukawa
coupling.  Except for parameter regions where the leptonic decays of
$\tz_2$ are strongly suppressed, $\tw_1\tz_2$ production leads to the
well known $3\ell$ ($=e,\mu$) signature for the Tevatron
collider\cite{trilep}.  As $\tan\beta$ increases beyond about 5, these
branchings again diverge, and increasingly $\tz_2\to\tau\bar{\tau}\tz_1$
dominates. Results of phenomenological analyses of trilepton signals for
$\tan\beta \sim 8-10$ obtained using older versions of ISAJET should,
therefore, be
interpreted with caution.  For $\tan\beta >40$, $\tz_2\to \tau\ttau_1$
opens up, and becomes quickly close to 100\%. Near the edge of parameter
space ($\tan\beta \sim 45$), the $\tz_2\to \tz_1 h$ decay opens up,
resulting in a reduction of the $\tz_2\to \tau\ttau_1$ branching
fraction.

In frame {\it c}), the large value of $m_0=500$ GeV yields a large value
of $m_{\ttau_1}$ (and other slepton masses) even if $\tan\beta$ is
large. In this case, the $\tw_1$ branching fractions are dominated by
the virtual $W$ boson, so that $B(\tw_1\to \tz_1 e\nu )$ and $B(\tw_1\to
\tz_1 \tau\nu )$ are nearly equal over almost the entire range of
$\tan\beta$. The branching fractions of $\tz_2$ for $m_0=500$ GeV are
shown in frame {\it d}). As in frame {\it c}), the branching fraction of
$\tz_2$ to $\tau$'s and $e$'s is nearly the same except when
$\tan\beta \geq 35-40$. In this case, there is a steadily increasing branching
fraction of $\tz_2\to \tz_1 b\bar{b}$ (and to some extent, also of 
$\tz_2 \to \tz_1 \tau\bar{\tau}$), which is mainly a reflection of
the increasing importance of virtual Higgs bosons in the $\tz_2$
three-body decays. We mention that for values of $\tan\beta$ somewhat below
the range where the decay $\tz_2 \to \tz_1 h$ becomes kinematically
allowed, contributions from {\it all} neutral Higgs bosons are important.

The above considerations motivated us to begin a systematic exploration
of how signals for supersymmetry may be altered if $\tan\beta$ indeed
turns out to be very large. To facilitate this analysis, we have 
incorporated the above calculations into the computer program 
ISAJET 7.32, so that realistic simulations of sparticle production and 
decay can be made for large $\tan\beta$.

Another important effect at large $\tan\beta$ is that 
tau Yukawa interactions can alter the mean polarization of the
$\tau$'s produced in chargino and neutralino decays. This, in turn, alters
the energy distribution of the visible decay products of the $\tau$. The 
$\tau$ polarization information is saved in ISAJET and used to dictate the
energy distribution of the $\tau$ decay products.

The rest of this paper is organized as follows. In Sec. II, we describe 
aspects of our event generation and analysis program for Tevatron experiments,
including a catalogue of some of the possible signals for supersymmetry
at large $\tan\beta$. In Sec. III, we present numerical results of our 
generation of supersymmetric signals and SM backgrounds, and show the reach of 
the Tevatron MI and TeV33 in the parameter space of the mSUGRA model.
In Sec. IV, we present a summary and conclusions from our work.
Some lengthy three-body decay formulae are included in the Appendix.

\section{Event simulation, signatures and cuts}

In several previous works\cite{bcpt}, a variety of signal 
channels for the discovery of
supersymmetry at the Tevatron were investigated, and plots were shown for
the reach of the Tevatron MI and TeV33 in the parameter space of the mSUGRA
model. The simulation of SUSY signal events was restricted to parameter
space values of $\tan\beta =2$ and 10. The promising discovery channels that
were investigated included the following:
\begin{itemize}
\item multi-jet $+\eslt$ events (veto hard, isolated leptons) (J0L),
\item events with a single isolated lepton plus jets $+\eslt$ (J1L),
\item events with two opposite sign isolated leptons plus jets $+\eslt$ (JOS),
\item events with two same sign isolated leptons plus jets $+\eslt$ (JSS),
\item events with three isolated leptons plus jets $+\eslt$ (J3L),
\item events with two isolated leptons $+\eslt$ (no jets, clean) (COS),
\item events with three isolated leptons $+\eslt$ (no jets, clean) (C3L).
\end{itemize}
In these samples, the number of leptons is {\it exactly} that indicated,
so that these samples are non-overlapping.
For Tevatron data samples on the order of 0.1 fb$^{-1}$, the J0L
signal generally gave the best reach for supersymmetry. It is the classic 
signature for detecting gluinos and squarks at hadron colliders. For 
larger data samples typical of those expected at the MI or TeV33, 
the C3L signal usually gave the best reach. In the present paper, we will
extend these results to the large $\tan\beta$ region of mSUGRA parameter
space; we will also look for new signatures which may be indicative of
supersymmetry at large $\tan\beta$.

By examining the branching fractions in Fig.~\ref{nfig2}, we expect in
general at large $\tan\beta$ that there would be a reduction in
supersymmetric events containing isolated $e$'s or $\mu$'s. We also
expect for large $\tan\beta$ and small $m_0$ a more conspicuous
presence of isolated $\tau$ leptons (defined by hadronic one- or three-
charged prong jets as discussed below). For large $\tan\beta$ and large
$m_0$, we expect an increased presence of tagged $b$-jets (defined by
displaced decay vertices or by identification of a muon inside of a
jet). For these reasons, we have expanded the set of event topologies
via which to search for SUSY to
include, in addition:
\begin{itemize}
\item multi-jet $+\eslt$ events which include at least one tagged $b$-jet
(J0LB),
\item multi-jet $+\eslt$ events which include at least one tagged $\tau$-jet
(J0LT),
\item multi-jet $+\eslt$ events which include at least 
either a tagged $b$-jet or
a tagged $\tau$-jet (J0LBT),
\item opposite-sign isolated dilepton plus jet $+\eslt$ events where 
at least one of
the isolated leptons is actually a tagged $\tau$-jet (JOST),
\item same-sign isolated dilepton plus jet $+\eslt$ events where at least one of
the isolated leptons is actually a tagged $\tau$-jet (JSST),
\item isolated trilepton plus jet $+\eslt$ events where at least one of
the isolated leptons is actually a tagged $\tau$-jet (J3LT),
\item clean opposite-sign isolated dilepton $+\eslt$ events where at least 
one of the isolated leptons is actually a tagged $\tau$-jet (COST),
\item clean isolated trilepton $+\eslt$ events where at least one of
the isolated leptons is actually a tagged $\tau$-jet (C3LT).
\end{itemize}
We note that some of these event samples are no longer non-overlapping;
for instance, the J0LB sample is a subset of the canonical $\eslt$ (J0L)
sample. In the tau samples, the lepton multiplicity is again exactly that 
indicated, except that at least one of the leptons is required to be
identified as a $\tau$.

To model the experimental conditions at the Tevatron, we use the toy
calorimeter simulation package ISAPLT. We simulate calorimetry covering
$-4<\eta <4$ with cell size $\Delta\eta\times\Delta\phi =0.1\times
0.0875$. We take the hadronic (electromagnetic) energy resolution to be
$70\% /\sqrt{E}$ ($15\% /\sqrt{E}$).  Jets are defined as hadronic
clusters with $E_T > 15$~GeV within a cone with $\Delta
R=\sqrt{\Delta\eta^2 +\Delta\phi^2} =0.7$. We require that $|\eta_j|
\leq 3.5$.  Muons and electrons are classified as isolated if they have
$p_T>5$~GeV, $|\eta (\ell )|<2.5$, and the visible activity within a
cone of $R=0.3$ about the lepton direction is less than $max(E_T(\ell
)/4,2\ {\rm GeV})$.  For tagged $b$-jets, we require a jet (using the
above jet requirement) to have in addition $|\eta_j|<2$ and to contain a
$b$-hadron. Then the jet is identified as a $b$-jet with a 50\%
efficiency. To identify a $\tau$-jet, we require a jet with just 1 or 3
charged prongs with $p_T>1$ GeV within $10^\circ$ of the jet axis, and
no other charged prongs within $30^\circ$ of the jet axis. The invariant
mass of the 3 prong jets must be less than $m_{\tau}$, and the net
charge of the 3 prongs should be $\pm 1$.  QCD jets with $p_T = 15$
($\geq 50$)~GeV are mis-identified as $\tau$ jets with a
probability~\cite{prob} of 0.5\% (0.1\%), with a linear interpolation in
between.  In our analysis, we neglect multiple scattering effects,
non-physics backgrounds from photon or jet misidentification, and make
no attempt to explicitly simulate any particular detector.

We incorporate in our analysis the following trigger conditions:
\begin{enumerate}
\item one isolated lepton with $p_T(\ell) > 15$~GeV and $\eslt >15$ GeV,
\item $\eslt >35$ GeV,
\item two isolated leptons each with $E_T>10$ GeV and $\eslt >10$ GeV,
\item one isolated lepton with $E_T>10$ GeV plus at least one jet plus
$\eslt >15$ GeV,
\item at least four jets per event, each with $E_T>15$ GeV.
\end{enumerate}
Thus, every signal or background event must satisfy at least one of the 
above conditions. 

We have generated the following physics background processes using ISAJET:
$t\bar t$ production, $W+$jets, $Z+$jets, $WW$, $WZ$ and $ZZ$ production
and QCD (mainly from $b\bar b$ and $c\bar c$ production). Each 
background subprocess was generated with subprocess final state particles
in $p_T$ bins of $25-50$ GeV, $50-100$ GeV, $100-200$ GeV, $200-400$ GeV
and $400-600$ GeV.

\section{The reach of the Fermilab Tevatron for mSUGRA}

We present our main results for the reach of the Tevatron for mSUGRA
at large $\tan\beta$ in the $m_0\ vs.\ m_{1/2}$ parameter space plane for
$A_0=0$ and for $\tan\beta =2,20,35$ and 45. Our results are shown for
$\mu >0$ only. For small $\tan\beta\sim 2$, the $\mu <0$ results differ
substantially from the $\mu >0$ results, and are shown in Ref. \cite{bcpt}.
As $\tan\beta$ increases, the positive and negative $\mu$ results become
increasingly indistinguishable.

In Fig.~\ref{nfig3} we show for orientation contours of constant
$m_{\tg}$ and $m_{\tq}$ in the $m_0\ vs.\ m_{1/2}$ plane.  The bricked
regions are excluded by either lack of appropriate electroweak symmetry
breaking, or due to the $\ttau_1$ or $\tw_1$ being the LSP instead of
the $\tz_1$. The gray regions are excluded by previous experimental
sparticle searches, and the excluded region~\cite{lep2} is dominantly
formed by the LEP2 bound that $m_{\tw_1}>80$ GeV~\cite{lepbnd}. The most
noticeable feature of Fig.~\ref{nfig3} is that the theoretically
excluded region increases significantly as $\tan\beta$ increases. In the
low $m_0$ region, this is due to the decrease in $\ttau_1$ mass, making
it become the LSP. The contours of $m_{\tg}$ and $m_{\tq}$ on the other
hand are relatively constant and change little with $\tan\beta$. The
region to the left of the dotted lines denotes where the decay modes
$\tw_1\to\ttau_1\nu$ and $\tz_2\to \ttau_1\tau$ become accessible.

As in our previous analysis of signals at low $\tan\beta$\cite{bcpt}, 
for channels involving jets, we require of all signals,
\begin{itemize}
\item jet multiplicity, $n_{jets}-n_{\tau -jets} \geq 2$,

\item $\etmiss > 40$~GeV, and

\item $E_T(j_1), \  E_T(j_2) \ > \ E_T^c$ and $\eslt > E_T^c$,

\end{itemize}
where the parameter $\ETC$ is taken to be
$E_T^c=15,40,60,80,100,120,140,160$ GeV. This requirement serves to give
some optimization of cuts for different masses of SUSY particles.
 
We generate signal events for each point on a 25~GeV~$\times$~25~GeV
grid in the $m_0-\mhf$ plane.  For an observable signal, we require at
least 5 signal events after all cuts (including those detailed below) are
imposed, with $N_{signal}$ exceeding $5\sqrt{N_{background}}$.  Any
signal is considered observable if it meets the observability criteria
for at least {\it one} of the values of $E_T^c$.  In addition, we
require the ratio of signal/background to exceed 0.2 for all
luminosities.

\subsection{Reach via the J0L channel}

As in Ref. \cite{bcpt}, for multijet$+\eslt$ events (J0L),
we require in addition to the above,
\begin{itemize}
\item transverse sphericity $S_T > 0.2$, and


\item  $\Delta\phi(\vec{\etmiss},\vec{E_{Tj}})>30^o$.

\end{itemize}

In Fig.~\ref{nfig4}, we show the Tevatron reach via the J0L channel.
Black squares denote parameter space points accessible to Tevatron
experiments with 0.1 fb$^{-1}$ of integrated luminosity (approximately
the Run I data sample); points denoted by gray squares are accessible
with 2 fb$^{-1}$ while those with open squares are accessible with 25
fb$^{-1}$. Points denoted by $\times$ are not visible at any of the
luminosity upgrade options considered. In frame {\it a}), no black
squares are visible; regions normally accessible to Tevatron experiments
with just 0.1
fb$^{-1}$ of integrated luminosity have been excluded by the negative
results of LEP2 searches for charginos. This is strictly valid only
within the model framework, and should not be regarded as a direct bound
on $m_{\tg}$. Regardless of the LEP2 bounds, Tevatron experiments should
directly probe this region via the independent search for strongly
interacting sparticles.  Note that even within the mSUGRA framework, for
$\mu <0$ and $\tan\beta =2$, where $m_{\tw_1}$ is considerable heavier
for the same $m_{1/2}$ values, there still exist parameter space points
accessible with only 0.1 fb$^{-1}$\cite{bcpt}.  A significant number of gray
squares appear in frame {\it a}), denoting regions with $m_{\tg}\sim
400$ GeV that can be probed at the MI. As $\tan\beta$ increases, the
theoretically excluded region absorbs some of these points at low $m_0$,
while some of the high $m_0$ points become inaccessible. In the latter
case, much of the signal actually comes from $\tw_1\overline{\tw_1}$ and
$\tw_1\tz_2$ production, and these particles decay decreasingly into
jetty final states, so the J0L signal diminishes. Finally, for very large
$\tan\beta =45$, none of the parameter space in this channel is open to
MI searches. For TeV33, we see that $m_{1/2}\sim 175$ GeV ($m_{\tg}\sim
475$ GeV) can be probed in all of the frames {\it a})-{\it d}) as long
as $m_0$ is not much larger. The
largest reach occurs when $E_T^c$ attains its largest value of
$E_T^c=160$ GeV.

\subsection{Reach via the J0LB channel}

In Fig.~\ref{nfig5}, we show the reach in the $\eslt +$jets channel,
where in addition we require at least one tagged $b$-jet
(J0LB). Comparing with Fig.~\ref{nfig4}, we see that the requirement of
a tagged $b$-jet considerably reduces the reach of the MI.  Furthermore,
the parameter space points with $m_{1/2}=175$ GeV are no longer
accessible to TeV33. In other words, a higher $E_T^c$ value is more
efficient in maximizing signal-to-background for large $m_{1/2}$ than
requiring an extra $b$-jet. However, for large $m_0$ and $m_{1/2}\sim
125-150$ GeV, the extra $b$-tag does somewhat increase the reach of
TeV33 for SUSY.  Comparison of Fig.~\ref{nfig4} and \ref{nfig5} shows
three additional points accessible in frame {\it a}), two in frame {\it
b}), and one in frame {\it d}). We have also tried to extend the
parameter space reach by requiring an identified $\tau$-jet (J0LT) or
either a $\tau$ or $b$ jet (J0LBT) along with $\eslt +$ jets. In both of
these cases, no additional reach was achieved beyond the results of Figs
\ref{nfig4} and \ref{nfig5}.

\subsection{Reach via the JOS and JSS channels}

The reach of Tevatron upgrades on the JOS channel is presented in
Fig.~\ref{nfig6}. 
We require, in addition to the conditions at the beginning of this Section,
\begin{itemize}
\item events with exactly two opposite sign isolated leptons ($e$ and $\mu$), 
with $E_T(\ell_1)> 10$~GeV and a veto of $\tau$-jets.
\end{itemize}
At the Tevatron at low $\tan\beta$, signals in this channel mainly come
from $\tw_1\tz_2$ production, where $\tz_2$ decays leptonically, and
$\tw_1$ decays hadronically, while top production is a major source of
SM background.  There is significant reach by the Tevatron
MI and TeV33 in this channel at low $\tan\beta$, as seen in frame {\it
a}). As $\tan\beta$ increases, the $\tz_2$ leptonic branching fraction
decreases (see Fig.~\ref{nfig2}), so that the MI has no reach in this
channel for $\tan\beta \geq 20$. The reach of TeV33 is severely limited in
this channel at high $\tan\beta$ as well.

We have also examined the reach of the MI and TeV33 for same-sign dileptons
(JSS channel), where we require in addition
\begin{itemize}
\item events to contain exactly two same sign
isolated leptons, again with $E_T(\ell_1)> 10$~GeV and a veto of
$\tau$-jets.
\end{itemize}
The reach of Tevatron upgrades in this channel for mSUGRA
is not very promising. The signal should result mainly from $\tg\tg$ and 
$\tg\tq$ production mechanisms, but these have only small cross  sections
for parameter space points beyond the reach of LEP2. We found almost no
reach for mSUGRA in this channel beyond the LEP2 bounds for {\it any}
values of $\tan\beta$.

We have also studied the Tevatron reach in the dilepton plus jets channels
where we required in addition that at least one of the leptons be a tagged 
$\tau$-jet: the JOST and JSST channels. In each of these cases,
a small increase in reach was obtained
for large values of $\tan\beta$ and low $m_0$ 
beyond the corresponding ``tau-less'' channels. Most of this additional
region can also be probed via the J3L channel discussed below,
so we do not show these results here.

\subsection{Reach via the J3L channel}

For small values of $\tan\beta$,
the J3L channel considerably increases the region of mSUGRA parameters
beyond what can be probed via the $\eslt$ channel at a high luminosity
Tevatron. In addition to the generic cuts for all the signals involving
jets, we require the following analysis cuts for the J3L channel:
\begin{itemize}
\item  events containing exactly three isolated leptons with 
$E_T(\ell_1)>10$~GeV and a veto of $\tau$-jets, plus
\item we veto events with $|M(\ell^+\ell^-)-M_Z|<8$~GeV.
\end{itemize}
The reach in the J3L channel
after all cuts are imposed is shown in Fig.~\ref{nfig7}.
Since the signal almost always
involves a leptonically decaying $\tz_2$, it is not surprising to see
that the large reach at low $\tan\beta$
is gradually diminished until there is almost no reach for $\tan\beta\sim 45$.

We have also examined the Tevatron reach in the trilepton plus jets
channels where we required in addition that at least one of the leptons
be a tagged $\tau$-jet: the J3LT channel.  As before, only a slight
additional reach was obtained at large $\tan\beta$ and low $m_0$ beyond
what could be probed via the ``tau-less'' J3L channel. Here, and in the
jetty dilepton channels mentioned above, this is presumably because
secondary leptons from tau decay tend to be soft, and fail to satisfy
the acceptance requirements.  Again, we do not show these results here.

\subsection{Reach via the C3L and C3LT channels}

For small $\tan\beta \sim 2$, and a large enough integrated luminosity,
the maximum reach of the Tevatron was often achieved via the clean
trilepton channel from $\tw_1\tz_2\to 3\ell+\eslt$.  For the C3L signal,
following our earlier analysis\cite{bcpt} we implement the following
cuts:
\begin{itemize}
\item we require 3 {\it isolated} leptons ($e$ and $\mu$)
within $|\eta_{\ell} |<2.5$
in each event, with $E_T(\ell_1)>20$
GeV, $E_T(\ell_2)>15$ GeV, and $E_T(\ell_3)>10$ GeV,
\item we require $\eslt >25$ GeV,
\item we require that the
invariant mass of any opposite-sign, same flavor dilepton pair not reconstruct
the $Z$ mass, {\it i.e.} we require that
$|m(\ell\bar{\ell})-M_Z|\geq 8$~GeV,
\item we finally require the events to be {\it clean}, {\it i.e.} we veto
events with jets.
\end{itemize}
Our calculated background in this channel is 0.2 fb.

In Fig.~\ref{nfig8}, we show the reach in the C3L channel for the four
cases of $\tan\beta$. In frame {\it a}), we see at low $\tan\beta$ that
indeed there is no reach beyond the current LEP2 bound in the C3L
channel for 0.1 fb$^{-1}$.  For the MI integrated luminosity, however,
there is considerable reach to values of $m_{1/2}\sim 225$ GeV, and for
TeV33, the reach extends to $m_{1/2}\sim 250$ GeV, corresponding to
$m_{\tg}\sim 700$ GeV!  As $\tan\beta$ increases, the branching fraction
for a
leptonic decay
of $\tz_2$ and $\tw_1$ decrease. In frame {\it b}), in fact, we
find {\it no} reach for SUSY via the C3L channel for MI and considerably
reduced reach for TeV33, except at large $m_0$.
For smaller values of $m_0$ a complicated interference between various
amplitudes reduces the leptonic decay width of $\tz_2$. As $\tan\beta$
increases even further to 35 and 45 as in frames {\it c}) and {\it d}),
the C3L reach is wiped out at low $m_0$.  Some reach remains at large
$m_0$ in frame {\it c}), where the branching fraction $BF(\tz_2\to
\ell\bar{\ell}\tz_1)\sim BF(Z\to \ell\bar{\ell})$.  In frame {\it d}),
most of this region also becomes
inaccessible because of the increased importance of (virtual) Higgs
boson mediated decays of $\tz_2$ which lead to a strong enhancement of
its decay to $b\bar{b}\tz_1$.

We have also examined the reach for clean trileptons, where one of the
leptons is actually an identified $\tau$-jet (C3LT).  In this case, we
relax the additional $p_T$ requirements on the leptons. This increases
the chance of detecting the softer secondary leptons from the decay of tau(s).
Trigger 4 presumably plays an important role for this
class of events.
The reach via this channel is shown in Fig.~\ref{nfig9}.
In frames {\it b}), {\it c}) and {\it d}), significant additional reach
is gained in the low $m_0$ regions, beyond that shown in any of the
previous figures! Notice that the region where the signal is observable
is where chargino and neutralino decays to real $\ttau_1$ are accessible
(see Fig.~\ref{nfig3}).  The reach in the C3LT channel effectively
extends the reach of TeV33 to $m_{1/2}\sim 250$ GeV for at least some
value of $m_0$ for all the values of large $\tan\beta$ considered.  We
remark that the gain in reach via channels involving taus is limited
because we require the presence of additional hard leptons ($e$ or
$\mu$), jets or $\eslt$ in order to be able to trigger on the
event. Because secondary leptons from the decay of a tau tend to be
soft, the development of an efficient $\tau$ trigger may significantly
enhance the reach when $\tan\beta$ is large.

\subsection{Reach via the COS and COST channels}

In our previous studies \cite{bcpt} we had already noted that for small
values of $\tan\beta$, a study of the
clean opposite sign dilepton channel (COS) would allow a confirmation of
the signal in the C3L channel for a large range of mSUGRA parameters.
For the COS channel, we require
\begin{itemize}
\item exactly two {\it isolated} OS (either $e$ or $\mu$ ) leptons
in each event, with $E_T(\ell_1)>10$ GeV and $E_T(\ell_2)>7$ GeV, and
$|\eta (\ell ) |<2.5$.
In addition, we require {\it no} jets, which
effectively reduces most of the $t\bar t$ background.
\item We require $\eslt >25$ GeV to remove
backgrounds from Drell-Yan dilepton production, and also
the bulk of the background from $\gamma^*, Z\to\tau\bar{\tau}$ decay.
\item We require $\phi (\ell\bar{\ell})<150^0$, to further reduce
$\gamma^*,Z\to\tau\bar{\tau}$ background.
\item We require the $Z$ mass cut:
invariant mass of any opposite-sign, same flavor dilepton pair not reconstruct
the $Z$ mass, {\it i.e.} $ \left| m(\ell\bar{\ell})- M_Z \right| > 8$ GeV.
Finally, we require $B=|\vec{\eslt}|+|p_T(\ell_1)|+|p_T(\ell_2)|<100$ GeV.
\end{itemize}
Our calculated background in this case is 64 fb.

We have checked that while there is an observable signal at the MI
(TeV33) for $m_{1/2}\sim 150$~(175)~GeV, and if $m_0 \alt 100$~GeV,
there is no observable signal for any of the allowed regions of the
plane if $\tan\beta \geq 20$.  We have also examined 
this channel by requiring in addition that at least one of the leptons 
be an identified $\tau$-jet (COST). In this case, no reach for mSUGRA was
found for any of the $\tan\beta$ values considered. We therefore do not 
show these figures.


\section{Summary and Conclusions}

To summarize the reach of Tevatron upgrades for large and small
$\tan\beta$, we show in Fig.~\ref{nfig11} the SUSY reach via all of the
channels that were examined, for both the upgrade options of the
Tevatron. Thus, if a parameter space point is accessible via any
channel, we place an appropriate box, corresponding to the integrated
luminosity that is required.  The cumulative reach shown in the figure
is completely established with
just four channels: J0L, J0LB, C3L and C3LT. For some points, the signal
may be observable in more than one of these or other channels 
studied in this paper.
It is possible that
some additional reach may be gained by combining several channels to gain a net
``$5\sigma$'' signal, even though the significance in each of these
channels is somewhat smaller.  We do not consider this added detail
here. 

We see from Fig.~\ref{nfig11} that as $\tan\beta$ increases, the SUSY
reach of Tevatron upgrades is significantly reduced. For the MI option,
there is no reach beyond current LEP2 bounds that can be established at
$\tan\beta =45$. The TeV33 option has some reach in all frames, but
clearly a much reduced reach for large $\tan\beta$. In particular, there
are parameter regions just beyond the current LEP2 bounds for which
there will be {\it no observable signal} even with the luminosity of
TeV33. 
The reduction of the reach is mostly due to the depletion of leptonic
signals, especially the clean three lepton signal, in the region of large
$\tan\beta$. Note that the branching ratio for $\tw_1$ and $\tz_2$
to decay into electrons and muons plus missing particles is actually quite
large if charginos and neutralinos dominantly decay into real or virtual
$\tilde{\tau}_1$. However, the secondary leptons produced in subsequent
$\tau$ decays are usually too soft to pass our trigger criteria or
acceptance cuts. It might be worthwhile to investigate whether these
cuts can be lowered without introducing unacceptably large backgrounds
({\it e.g.} from heavy flavors, where the lepton happens to be isolated
and the jet is lost, or from jets faking leptons) or via a development of a
special trilepton trigger.

Modes with identified (hadronically decaying) taus could only partly
compensate this loss of reach in the leptonic channels.  Again the
problem seems to be that the hadronic decay products of the $\tau$
leptons are frequently too soft to pass the cut $E_T(\tau-{\rm jet})>15$
GeV. It might be worthwhile to study if this cut can be lowered, {\it
e.g.} by focussing only on one--prong $\tau$ decays, for which QCD
backgrounds are much smaller than in the three--prong channel. In
addition, the triggers adopted in our study are not very efficient for
events with rather soft leptons plus $\tau-$jets, as in our C3LT sample.
We therefore believe that the reach of future Tevatron runs could be
extended significantly in the region of large $\tan \beta$ if it is
possible to devise strategies to reliably identify, and perhaps even
trigger on taus with visible $p_T$ smaller than 15~GeV.  We remark,
however, that even without such
developments, experiments at the LHC will probe the entire parameter
plane shown at least via the $\eslt$ channel.

%
\acknowledgments
We thank Vernon Barger for reading the manuscript.
One of us (XT) is grateful for the hospitality of 
the Asia-Pacific Centre for Theoretical
Physics where part of this work was
carried out. HB and XT thank the Aspen Center for Physics for 
hospitality during the period that part of this work was
done.
This research was supported in part by the U.~S. Department of Energy
under contract number DE-FG05-87ER40319, DE-AC02-76CH00016, and
DE-FG-03-94ER40833. 
\bigskip
\appendix
\centerline{\bf Appendix: Sparticle Decay Widths for Large $\tan\beta$}
\bigskip

In this Appendix we give analytical expressions for those three--body partial
widths that are sensitive to $b$ or $\tau$ Yukawa couplings and/or to
$\tilde{f}_L - \tilde{f}_R$ mixing ($f=b, \tau$). We first list the
relevant couplings, and then give results for 
$\wt{Z}_i \rightarrow \wt{Z}_j f \bar{f}$, 
$\wt{W}_i \rightarrow \tau \nu_{\tau} \wt{Z}_j$,
$\wt{Z}_i \rightarrow \wt{W}_j f \bar{f}$, and 
$\tilde{g} \rightarrow b t \wt{W}_i$.

Many of the couplings and kinematic functions
that enter our computations have been defined in
our earlier papers. Instead of rewriting these lengthy definitions
again, we provide the reader with references to the papers from which
these couplings are used. In these studies, the two charginos were
denoted by $\tw_-$ and $\tw_+$ instead of $\tw_1$ and $\tw_2$,
respectively. Also, the lighter (heavier) neutral $CP$ even Higgs
scalar was denoted by $H_{l}$ ($H_h$) rather than by $h$ ($H$), while
the $CP$ odd pseudoscalar was denoted by $H_p$ rather than $A$. 
The corresponding couplings are characterized by superscripts $l$, $h$
and $p$. To facilitate the use of these couplings from the earlier
literature, we use this older notation to denote the charginos and
neutral Higgs bosons in the formulae listed in this
Appendix. 

\subsection{Couplings}

The couplings of electroweak neutralinos and charginos to a fermion and
a sfermion are affected by mixing between $SU(2)$ doublet ($L-$type)
and singlet ($R-$type) sfermions. We write the sfermion mass eigenstates as:
\beq \label{ea1}
\tilde{f}_1 &= \cos \! \theta_f \tilde{f}_L - \sin \! \theta_f \tilde{f}_R;
\nonumber \\
\tilde{f}_2 &= \sin \! \theta_f \tilde{f}_L + \cos \! \theta_f \tilde{f}_R,
\eeq
where $\tilde{f}_1$ denotes the lighter eigenstate. Since there is no
$L-R$ mixing in the sneutrino sector, some couplings remain unaffected. We
list these for completeness, using the notation of Refs.\cite{bbkt} and
\cite{btw}:
\ben 
\label{ea2} 
\beq
\wt{A}_{\wt{Z}_i}^{\nu} & = ( g v_3^{(i)} - g' v_4^{(i)} )/\sqrt{2};
\label{ea2a} \\
\wt{A}_{\wt{W}_-}^{\tau} &= -g \sin \gamma_R;
\label{ea2b} \\
\wt{A}_{\wt{W}_-}^{\nu} &= - g \sin \gamma_L ; \label{ea2c} \\
B_{\wt{W}_-}^{\tau} &= -f_{\tau} \cos \gamma_L \label{ea2d}.
\eeq 
\een
Here, $g$ and $g'$ are the $SU(2)$ and $U(1)_Y$ gauge couplings, and $f_f$
the Yukawa couplings of fermion $f$.
The corresponding couplings of the heavier chargino mass eigenstate
$\wt{W}_+$ can be obtained by the substitutions \cite{bbkt}
\be \label{ea3}
\wt{W}_- \rightarrow \wt{W}_+: \ \ \cos \gamma_{L,R} \rightarrow - 
\theta_{x,y} \sin \gamma_{L,R}; \ \ \ 
\sin \gamma_{L,R} \rightarrow \theta_{x,y} \cos \gamma_{L,R}.
\ee

In the calculation of the partial widths, we will ignore terms $\propto
m_b, m_\tau$ when doing the Dirac traces. It then becomes convenient to
write the matrix elements in terms of couplings to fermions with fixed
chirality. In the following we denote all left--handed couplings with
the symbol $\alpha$, and right--handed couplings with $\beta$. The
chargino couplings to the lighter third generation squark mass eigenstates 
can be written as:
\ben \label{ea4} \beq
\alpha_{\wt{W}_-}^{\tilde{t}_1} &= -g \sin \gamma_R \cos \theta_t +
f_t \cos \gamma_R \sin \theta_t ; \label{ea4a} \\
\beta_{\wt{W}_-}^{\tilde{t}_1} &= - f_b \cos \gamma_L \cos \theta_t;
\label{ea4b} \\
\alpha_{\wt{W}_-}^{\tilde{b}_1} &= -g \sin \gamma_L \cos \theta_b
+ f_b \cos \gamma_L \sin \theta_b ; \label{ea4c} \\
\beta_{\wt{W}_-}^{\tilde{b}_1} &= -f_t \cos \gamma_R \cos \theta_b.
\label{ea4d}
\eeq \een
The corresponding couplings to third generation sleptons can be obtained
by the substitutions:
\be \label{ea5}
\tilde{q} \rightarrow \tilde{l}: \ \ \ \theta_t \rightarrow 0; \ \ \
\theta_b \rightarrow \theta_\tau; \ \ \ f_t \rightarrow 0; \ \ \ 
f_b \rightarrow f_\tau.
\ee
Similarly, the couplings to the heavier sfermion mass eigenstates 
$\tilde{f}_2$ can be obtained by substituting:
\be \label{ea6}
\tilde{f}_1 \rightarrow \tilde{f}_2: \ \ \ \cos \theta_f \rightarrow
\sin \theta_f; \ \ \ \sin \theta_f \rightarrow - \cos \theta_f.
\ee
Finally, the couplings of the heavier chargino state can again be
computed using Eq.(\ref{ea3}).

The couplings of neutralinos to $b$ and $\tau$ (s)fermions can be written
as:
\ben \label{ea7} \beq
\alpha_{\wt{Z}_i}^{\tilde{f}_1} &= \wt{A}_{\wt{Z}_i}^f \cos \theta_f
- f_f v_2^{(i)} \sin \theta_f; \label{ea7a} \\
\beta_{\wt{Z}_i}^{\tilde{f}_1} &= f_f v_2^{(i)} \cos \theta_f
+ \wt{B}_{\wt{Z}_i}^f \sin \theta_f, \label{ea7b}
\eeq \een
where $\wt{A}_{\wt{Z}_i}^f, \ \wt{B}_{\wt{Z}_i}^f$ are as in Ref.\cite{btw}.
The couplings to fermions with weak isospin $I_3 = +1/2$ can be computed
from eqs.(\ref{ea7}) by inserting the corresponding unmixed couplings;
in addition, one has to replace the component $v_2^{(i)}$ of the neutralino
eigenvector by $v_1^{(i)}$. The couplings to heavier sfermion eigenstates
can again be obtained by applying Eq.(\ref{ea6}).

Finally, we introduce the charged Higgs -- chargino -- neutralino
couplings
\ben \label{ea8} \beq
\alpha^{(i)}_{\wt{W}_-} &= \cosb A_2^{(i)}, \ \ \
\beta^{(i)}_{\wt{W}_-} = - \sinb A_4^{(i)}; \label{ea8a} \\
\alpha^{(i)}_{\wt{W}_+} &= \cosb A_1^{(i)}\theta_y, \ \ \
\beta^{(i)}_{\wt{W}_+} = - \sinb A_3^{(i)}\theta_x, \label{ea8b}
\eeq \een
where $i$ is the neutralino index; 
the $A_k^{(i)}$ can be found in Ref.\cite{bbkmt}.

\subsection{$\wt{Z}_j \rightarrow \wt{Z}_i f \bar{f}$ Decays}

We are now in a position to present our results for the partial widths
for decays involving third generation fermions. We begin with the decay
of a neutralino into a lighter neutralino and a $b \bar{b}$ or
$\tau^+ \tau^-$ pair. This decay can proceed through the exchange of
the two sfermion mass eigenstates $\tilde{f}_{1,2}$, through the exchange
of a $Z$ boson, or through the exchange of one of the three neutral
Higgs bosons of the MSSM. The partial width can therefore be written as
\be \label{ea9}
\Gamma(\wt{Z}_j \rightarrow \wt{Z}_i f \bar{f}) = \frac{1}{2}
N_c(f) \frac{1} {(2 \pi)^5} \frac{1} {2 m_{\wt{Z}_j}} \left(
\Gamma_{\tilde{f}} + \Gamma_Z + \Gamma_{H_{l,h}} + \Gamma_{H_p} +
\Gamma_{Z \tilde{f}} + \Gamma_{H_{l,h} \tilde{f}} 
+ \Gamma_{H_p \tilde{f}} \right),
\ee
where the color factor $N_c(f) = 3$ (1) for $f = b \ (\tau)$.
Recall that we set $m_f=0$ when evaluating Dirac traces. As a result, 
the Higgs and $Z$ exchange diagrams do not interfere with each 
other.\footnote{This would be a very bad approximation for
$\wt{Z}_j \rightarrow \wt{Z}_i t \bar{t}$ decays. However, if these
decays are allowed, $\wt{Z}_j$ has numerous 2--body decay modes into
real gauge and Higgs bosons and lighter neutralinos and charginos.
The branching ratios for neutralino 3--body decays into top quarks are 
therefore always negligibly small. Analogous remarks apply to
$\wt{W}_j \rightarrow \wt{Z}_i t \bar{b}$ decays.}

The {\em pure sfermion exchange contribution} is given by
\be \label{ea10}
\Gamma_{\tilde f} = \Gamma_{\tilde{f}_1} + \Gamma_{\tilde{f}_2} + 
\Gamma_{\tilde{f}_{1,2}},
\ee
where
\ben \label{ea11} \beq
\Gamma_{\tilde{f}_k} &= \Gamma_{LL}^{\tilde{f}_k} +
\Gamma_{RR}^{\tilde{f}_k} + \Gamma_{LR}^{\tilde{f}_k} \ \ \ \ (k=1,2);
\label{ea11a} \\
\Gamma_{\tilde{f}_{1,2}} &= 
\Gamma_{L}^{\tilde{f}_1} \Gamma_{L}^{\tilde{f}_2} +
\Gamma_{L}^{\tilde{f}_1} \Gamma_{R}^{\tilde{f}_2} +
\Gamma_{R}^{\tilde{f}_1} \Gamma_{L}^{\tilde{f}_2} +
\Gamma_{R}^{\tilde{f}_1} \Gamma_{R}^{\tilde{f}_2}. \label{ea11b}
\eeq \een
Here, the subscripts $L$ and $R$ refer to the chirality of the SM fermion
coupling to the heavier neutralino $\wt{Z}_j$. The quantities
appearing in Eq.(\ref{ea11}) are:
\ben \label{ea12} \beq
\Gamma_{LL}^{\tilde{f}_k} &= 4 \left( \alpha_{\wt{Z}_j}^{\tilde{f}_k}
\right)^2 \left\{ \left[ \left( \alpha_{\wt{Z}_i}^{\tilde{f}_k}
\right)^2 + \left( \beta_{\wt{Z}_i}^{\tilde{f}_k} \right)^2 \right]
\psi( m_{\wt{Z}_j}, m_{\tilde{f}_k}, m_{\wt{Z}_i} ) \right. \nonumber \\
& \left. \hspace*{22mm} + (-1)^{\theta_i + \theta_j} 
\left( \alpha_{\wt{Z}_i}^{\tilde{f}_k} \right)^2 
\phi( m_{\wt{Z}_j}, m_{\tilde{f}_k}, m_{\wt{Z}_i} ) \right\}; 
\label{ea12a} \\
\Gamma_{RR}^{\tilde{f}_k} &= 4 \left( \beta_{\wt{Z}_j}^{\tilde{f}_k}
\right)^2 \left\{ \left[ \left( \alpha_{\wt{Z}_i}^{\tilde{f}_k}
\right)^2 + \left( \beta_{\wt{Z}_i}^{\tilde{f}_k} \right)^2 \right]
\psi( m_{\wt{Z}_j}, m_{\tilde{f}_k}, m_{\wt{Z}_i} ) \right. \nonumber \\
& \left. \hspace*{22mm} + (-1)^{\theta_i + \theta_j} 
\left( \beta_{\wt{Z}_i}^{\tilde{f}_k} \right)^2 
\phi( m_{\wt{Z}_j}, m_{\tilde{f}_k}, m_{\wt{Z}_i} ) \right\}; 
\label{ea12b} \\
\Gamma_{LR}^{\tilde{f}_k} &= -8 \alpha_{\wt{Z}_j}^{\tilde{f}_k}
\beta_{\wt{Z}_j}^{\tilde{f}_k}  \alpha_{\wt{Z}_i}^{\tilde{f}_k}
\beta_{\wt{Z}_i}^{\tilde{f}_k} 
 Y(m_{\wt{Z}_j}, m_{\tilde{f}_k}, m_{\tilde{f}_k}, m_{\wt{Z}_i} ); 
\label{ea12c} \\
\Gamma_L^{\tilde{f}_1} \Gamma_L^{\tilde{f}_2} &= 8
\alpha_{\wt{Z}_j}^{\tilde{f}_1} \alpha_{\wt{Z}_j}^{\tilde{f}_2} \left\{
\left[ \alpha_{\wt{Z}_i}^{\tilde{f}_1}
       \alpha_{\wt{Z}_i}^{\tilde{f}_2} +
       \beta_{\wt{Z}_i}^{\tilde{f}_1}
       \beta_{\wt{Z}_i}^{\tilde{f}_2} \right]
\wt{\psi}(m_{\wt{Z}_j}, m_{\tilde{f}_1}, m_{\tilde{f}_2}, m_{\wt{Z}_i} )
\right. \nonumber \\ & \left. \hspace*{22mm}
 + (-1)^{\theta_i + \theta_j}  \alpha_{\wt{Z}_i}^{\tilde{f}_1}
\alpha_{\wt{Z}_i}^{\tilde{f}_2} 
\tilde{\phi}(m_{\wt{Z}_j}, m_{\tilde{f}_1}, m_{\tilde{f}_2}, m_{\wt{Z}_i} )
\right\} ; \label{ea12d} \\
\Gamma_R^{\tilde{f}_1} \Gamma_R^{\tilde{f}_2} &= 8
\beta_{\wt{Z}_j}^{\tilde{f}_1} \beta_{\wt{Z}_j}^{\tilde{f}_2} \left\{
\left[ \alpha_{\wt{Z}_i}^{\tilde{f}_1}
       \alpha_{\wt{Z}_i}^{\tilde{f}_2} +
       \beta_{\wt{Z}_i}^{\tilde{f}_1}
       \beta_{\wt{Z}_i}^{\tilde{f}_2} \right]
\wt{\psi}(m_{\wt{Z}_j}, m_{\tilde{f}_1}, m_{\tilde{f}_2}, m_{\wt{Z}_i} )
\right. \nonumber \\ & \left. \hspace*{22mm}
 + (-1)^{\theta_i + \theta_j}  \beta_{\wt{Z}_i}^{\tilde{f}_1}
\beta_{\wt{Z}_i}^{\tilde{f}_2} 
\tilde{\phi}(m_{\wt{Z}_j}, m_{\tilde{f}_1}, m_{\tilde{f}_2}, m_{\wt{Z}_i} )
\right\}; \label{ea12e} \\
\Gamma_{L}^{\tilde{f}_1} \Gamma_{R}^{\tilde{f}_2}
 &= -8 \alpha_{\wt{Z}_j}^{\tilde{f}_1}
\beta_{\wt{Z}_j}^{\tilde{f}_2}  \alpha_{\wt{Z}_i}^{\tilde{f}_2}
\beta_{\wt{Z}_i}^{\tilde{f}_1} 
 Y(m_{\wt{Z}_j}, m_{\tilde{f}_1}, m_{\tilde{f}_2}, m_{\wt{Z}_i} ); 
\label{ea12f} \\
\Gamma_{L}^{\tilde{f}_2} \Gamma_{R}^{\tilde{f}_1}
 &= -8 \alpha_{\wt{Z}_j}^{\tilde{f}_2}
\beta_{\wt{Z}_j}^{\tilde{f}_1}  \alpha_{\wt{Z}_i}^{\tilde{f}_1}
\beta_{\wt{Z}_i}^{\tilde{f}_2} 
 Y(m_{\wt{Z}_j}, m_{\tilde{f}_1}, m_{\tilde{f}_2}, m_{\wt{Z}_i} ).
\label{ea12g} 
\eeq \een
The kinematic functions $\psi, \ \phi,$ and $Y$ are given in
Ref.\cite{btw2}\footnote{Note that the third line in Eq.(A6h) of that
paper should come with a positive overall sign. Furthermore, the last
term in the first denominator in Eq.(A6a) should be $m^2_{\tilde t}$,
rather than $m_t^2$. Of course, $m_t$ is repaced by the appropriate
fermion mass in the definition of these functions. Finally, although the
number of arguments appearing in the $Y$ function are different from
that in Ref.\cite{btw2}, the correspondence is obvious.}, and $\theta_i$
is 0 (1) if the sign of the $i^{\rm th}$ eigenvalue of the neutralino mass
matrix is positive (negative) \cite{bbkt}. The functions $\tilde{\psi}$
and $\tilde{\phi}$, which depend on two sfermion masses are
generalizations of the functions $\psi$ and $\phi$ which depend on just
one sfermion mass: to define $\tilde{\psi}$, we simply split the squared
factor where the stop mass occurs in Eq.~(A6a) of Ref.\cite{btw2}, into
two such factors, with each one containing a different sfermion mass. 
Similarly, $\tilde{\phi}$ is generalized from $\phi$ by writing
$m_{\tf_1}$ in the first factor outside the square parenthesis in
Eq.~(A6b) of Ref.\cite{btw2}, and $m_{\tf_2}$ inside the square
parenthesis. In other words, when the two sfermions $\tf_1$ and $\tf_2$
have the same mass, $\tilde{\psi} = \psi$ and $\tilde{\phi} = \phi$.

For completeness, we also give the {\em squared $Z$ exchange contribution},
which is not affected by sfermion mixing:
\beq \label{ea13}
\Gamma_Z &= 128 e^2 |W_{ij}|^2 \left( \alpha_f^2 + \beta_f^2 \right)
m_{\wt{Z}_j} \frac {\pi^2}{2} \int_{m_{\wt{Z}_i}}^{E_{\rm max}}
 dE \frac { B_f \sqrt{ E^2 - m^2_{\wt{Z}_i} } }
{ \left( m^2_{\wt{Z}_i} + m^2_{\wt{Z}_j} - M_Z^2 - 2 E m_{\wt{Z}_j} \right)^2 }
\nonumber
 \\
&\cdot \left\{ E \left[ m^2_{\wt{Z}_i} + m^2_{\wt{Z}_j} -
(-1)^{\theta_i+\theta_j} 2 m_{\wt{Z}_i} m_{\wt{Z}_j} \right]
- m_{\wt{Z}_j} \left( E^2 + m^2_{\wt{Z}_i}+\frac{B_f}{3}(E^2-m^2_{\wt{Z}_i})
\right)\right.\nonumber\\ 
&\qquad\left.+ (-1)^{\theta_i + \theta_j} m_{\wt{Z}_i} 
\left( m^2_{\wt{Z}_i} + m^2_{\wt{Z}_j}-2m_f^2 \right) \right\}.
\eeq
Here, $e$ is the QED coupling, $W_{ij}$ is the $Z \wt{Z}_i \wt{Z}_j$
coupling given in Ref.\cite{dkt}, and $\alpha_f, \ \beta_f$ are the
left-- and right--handed $Z f \bar{f}$ couplings in the notation of
Ref.\cite{dt}. Finally, the upper integration limit is given by
\be \label{ea14}
E_{\rm max} = \frac { m^2_{\wt{Z}_i} + m^2_{\wt{Z}_j}-4m_f^2 } 
{2 m_{\wt{Z}_j}} 
\ee
and
\be
B_f=\sqrt{1-\frac{4m_f^2}{m^2_{\wt{Z}_i} + m^2_{\wt{Z}_j}-2E m_{\wt{Z}_j}}}.
\ee

The {\em pure scalar Higgs exchange contribution} can also be written as
a single integral:
\beq \label{ea15}
\Gamma_{H_{l,h}} &= 2 \pi^2 \left( \frac {g m_f} {M_W \cos \! \beta}
\right)^2 m_{\wt{Z}_j} \int_{m_{\wt{Z}_i}}^{E_{\rm max}} dE B_f
\sqrt{E^2 - m^2_{\wt{Z}_i}} \nonumber \\
& \hspace*{15mm}
 \cdot \left( m^2_{\wt{Z}_i} + m^2_{\wt{Z}_j} - 2 m_{\wt{Z}_j} E-2m_f^2 \right)
\left[ E + (-1)^{\theta_i + \theta_j} m_{\wt{Z}_i} \right] \nonumber \\
& \hspace*{15mm}
 \cdot \left[ \frac {\sin\alpha \left( X^l_{ij} + X^l_{ji} \right) }
{ m^2_{\wt{Z}_i} + m^2_{\wt{Z}_j} - 2 m_{\wt{Z}_j} E - m^2_{H_l} } +
 \frac {\cos\alpha \left( X^h_{ij} + X^h_{ji} \right) }
{ m^2_{\wt{Z}_i} + m^2_{\wt{Z}_j} - 2 m_{\wt{Z}_j} E - m^2_{H_h} } \right]^2.
\eeq
Here, $X_{ij}^{l,h}$ are the couplings of the light and heavy neutral
scalar Higgs boson to two neutralinos and
$\alpha$ is the angle describing mixing in the scalar Higgs sector as
defined in Ref.\cite{bbkmt}, and
$m^2_{H_{l.h}}$ are the masses of the two Higgs bosons. The upper integration
limit is again given by Eq.(\ref{ea14}).

The {\em squared pseudoscalar Higgs exchange contribution} can be cast
in a quite similar form:
\beq \label{ea16}
\Gamma_{H_p} &= 2 \pi^2 \left[ \frac {g m_f \tan \beta} {M_W} \left(
X^p_{ij} + X^p_{ji} \right) 
\right]^2 m_{\wt{Z}_j} \nonumber \\ & \cdot
\int_{m_{\wt{Z}_i}}^{E_{\rm max}} dE
B_f \sqrt{E^2 - m^2_{\wt{Z}_i}} 
\frac{ \left( m^2_{\wt{Z}_i} + m^2_{\wt{Z}_j} - 2 m_{\wt{Z}_j} E-2m_f^2 \right)
\left[ E - (-1)^{\theta_i + \theta_j} m_{\wt{Z}_i} \right] }
{ \left( m^2_{\wt{Z}_i} + m^2_{\wt{Z}_j} - 2 m_{\wt{Z}_j} E - m^2_{H_p}
\right)^2 }.
\eeq
The couplings $X^p_{ij}$ can again be found in Ref.\cite{bbkmt}.

We now turn to the various interference terms listed in Eq.(\ref{ea9}).
The {\em $Z-$sfermion interference contributions} can be written as
\be \label{ea17}
\Gamma_{Z \tilde{f}} = \Gamma_{Z \tilde{f}_1} + \Gamma_{Z \tilde{f}_2},
\ee
with
\beq \label{ea18}
\Gamma_{Z \tilde{f}_k} &= 32 e \wt{W}_{ij} \left[ 
\alpha_{\wt{Z}_i}^{\tilde{f}_k} \alpha_{\wt{Z}_j}^{\tilde{f}_k} 
\left( \alpha_f - \beta_f \right) -
\beta_{\wt{Z}_i}^{\tilde{f}_k} \beta_{\wt{Z}_j}^{\tilde{f}_k} 
\left( \alpha_f + \beta_f \right) \right] \frac {\pi^2} {2 m_{\wt{Z}_j}}
\nonumber \\
& \cdot \int_{4m_f^2}^{\left( m_{\wt{Z}_j} - m_{\wt{Z}_i} \right)^2}
\frac {ds} {s-M_Z^2} \left\{ - \frac{1}{2} Q' \left( m_{\wt{Z}_j} E_Q 
+ m^2_{\tilde{f}_k} - m^2_{\wt{Z}_j} - s-m_f^2 \right)
\right. \nonumber \\ 
& \hspace*{2mm}
- \frac{1} {4 m_{\wt{Z}_j}} \left[ \left( m^2_{\tilde{f}_k}-m^2_{\wt{Z}_i}-m_f^2
\right) \left( m^2_{\tilde{f}_k} - m^2_{\wt{Z}_j}-m_f^2 \right) 
+ (-1)^{\theta_i + \theta_j} m_{\wt{Z}_i} m_{\wt{Z}_j} (s-2m_f^2) \right]
\nonumber \\
&\hspace*{2mm} \left. \cdot\log 
\frac { m_{\wt{Z}_j} \left( E_Q + Q' \right) - \mu^2 }
{ m_{\wt{Z}_j} \left( E_Q - Q' \right) - \mu^2 } \right\}.
\eeq
Here we have introduced the quantities
\be \label{ea19}
\mu^2 = s + m^2_{\tilde{f}_k} - m^2_{\wt{Z}_i}-m_f^2, \ \ \ 
E_Q = \frac {s + m^2_{\wt{Z}_j} - m^2_{\wt{Z}_i} } {2 m_{\wt{Z}_j} }, \ \ \
Q = \sqrt{E_Q^2 - s},
\ee
and
\beq
Q'=Q\sqrt{1-\frac{4m_f^2}{s}} .\nonumber
\eeq
The real coupling $\wt{W}_{mn}$ is defined to be,
\be \label{ea20}
\wt{W}_{mn} = (-i)^{\theta_m + \theta_n} (-1)^{\theta_m} W_{mn},
\ee
with $W_{mn}$ given in Ref.\cite{dkt}.

Finally, the {\em Higgs--sfermion interference contributions} can be
written as
\be \label{ea21}
\Gamma_{H_{l,h,p} \tilde{f}} = \Gamma_{H_{l,h,p} \tilde{f}_1}
+ \Gamma_{H_{l,h,p} \tilde{f}_2},
\ee
where $H_l$, $H_h$ and $H_p$ again denote the light scalar, heavy scalar,
and pseudoscalar Higgs boson, respectively. The separate contributions
in Eq.~(\ref{ea21}) are given by:
\ben \label{ea22} \beq
\Gamma_{H_l \tilde{f}_k} & = \frac {2 \pi^2} {m_{\wt{Z}_j}}
\frac { g m_f \sin \alpha} {M_W \cosb} \left( X^l_{ji} + X^l_{ij} \right)
\left[ \alpha_{\wt{Z}_i}^{\tilde{f}_k} \beta_{\wt{Z}_j}^{\tilde{f}_k} 
     + \alpha_{\wt{Z}_j}^{\tilde{f}_k} \beta_{\wt{Z}_i}^{\tilde{f}_k}   
\right] \cdot (-1)^{\theta_i+\theta_j}
\nonumber \\ 
& \cdot J_H(m_{\wt{Z}_j}, m_{\tilde{f}_k}, m_{H_l}, m_{\wt{Z}_i},
               \theta_i + \theta_j) ; 
\label{ea22a} \\
\Gamma_{H_h \tilde{f}_k} & = \frac {2 \pi^2} {m_{\wt{Z}_j}}
\frac { g m_f \cos \alpha} {M_W \cosb} \left( X^h_{ji} + X^h_{ij} \right)
\left[ \alpha_{\wt{Z}_i}^{\tilde{f}_k} \beta_{\wt{Z}_j}^{\tilde{f}_k} 
     + \alpha_{\wt{Z}_j}^{\tilde{f}_k} \beta_{\wt{Z}_i}^{\tilde{f}_k}   
\right] \cdot (-1)^{\theta_i+\theta_j}
\nonumber \\ 
& \cdot J_H(m_{\wt{Z}_j}, m_{\tilde{f}_k}, m_{H_h}, m_{\wt{Z}_i},
               \theta_i + \theta_j) ; 
\label{ea22b} \\
\Gamma_{H_p \tilde{f}_k} & = \frac {2 \pi^2} {m_{\wt{Z}_j}}
\frac { g m_f \tanb} {M_W} \left( X^p_{ji} + X^p_{ij} \right)
\left[ \alpha_{\wt{Z}_i}^{\tilde{f}_k} \beta_{\wt{Z}_j}^{\tilde{f}_k} 
     + \alpha_{\wt{Z}_j}^{\tilde{f}_k} \beta_{\wt{Z}_i}^{\tilde{f}_k}   
\right] \cdot (-1)^{1+\theta_i+\theta_j}
\nonumber \\ 
& \cdot J_H(m_{\wt{Z}_j}, m_{\tilde{f}_k}, m_{H_p}, m_{\wt{Z}_i},
               1 + \theta_i + \theta_j) ; 
\label{ea22c}
\eeq \een
The function $J_H$ is defined as
\beq \label{ea23}
J_H &(m_{\wt{Z}_j}, m_{\tilde{f}}, m_{H}, m_{\wt{Z}_i}, \theta) =
\int_{4m_f^2}^{\left(m_{\wt{Z}_j} - m_{\wt{Z}_i} \right)^2} \frac {ds} {s-m_H^2}
\\ 
&\cdot \left[ \frac{1}{2} s Q' + \frac {s m^2_{\tilde f}-m_f^2(m_{\wt{Z}_i^2}+
m_{\wt{Z}_j^2}) + (-1)^\theta
m_{\wt{Z}_i} m_{\wt{Z}_j}(s-2m_f^2) } {4 m_{\wt{Z}_j} } 
\cdot \log \frac { m_{\wt{Z}_j} \left( E_Q + Q' \right) - \mu^2 }
{ m_{\wt{Z}_j} \left( E_Q - Q' \right) - \mu^2 } \right],\nonumber
\eeq
where $\mu^2, \ E_Q$, $Q$ and $Q'$ have been defined in Eq.(\ref{ea19}).
 
\subsection{$\wt{W}_j \rightarrow \wt{Z}_i \tau \nu_\tau$ Decays}

These decays proceed via the exchange of a $W$ boson, a charged or neutral
third generation slepton, or a charged Higgs boson. The partial widths can 
thus be written as
\be \label{ea24}
\Gamma(\wt{W}^-_j \rightarrow \wt{Z}_i \tau^- \bar{\nu}_{\tau}) = 
\frac {1}{2} \frac{1}{(2 \pi)^5} \frac {1} {2 m_{\wt{W}_j}} \left(
\Gamma_W + \Gamma_{\tilde \nu} + \Gamma_{\tilde \tau} + \Gamma_H +
\Gamma_{W \tilde{\nu}} + \Gamma_{W \tilde{\tau}} + 
\Gamma_{\tilde{\nu} \tilde{\tau}} + \Gamma_{H \tilde{\nu}} +
\Gamma_{H \tilde{\tau}} \right).
\ee
The Higgs and $W$ exchange contributions do not interfere, since we
neglected terms $\propto m_{\tau}$ when doing the Dirac algebra.

The {\em squared $W$ exchange contribution} is given by
\beq \label{ea25}
\Gamma_W = 4 g^4 \frac {\pi^2} {3} & \int_{m_{\wt{Z}_i}}^{E_{\rm max}}
dE \frac { \sqrt {E^2 - m^2_{\wt{Z}_i}} }
{ \left( m^2_{\wt{W}_j} + m^2_{\wt{Z}_i} - 2 m_{\wt{W}_j} E - M_W^2
 \right)^2 } \nonumber \\ 
& \cdot \left\{  \left( \left| X^i_j \right|^2 + \left| Y^i_j \right|^2
\right) 
\left[ 3 \left( m^2_{\wt{W}_j} + m^2_{\wt{Z}_i} \right) m_{\wt{W}_j} E
           - 2 m^2_{\wt{W}_j} \left( 2 E^2 + m^2_{\wt{Z}_i} \right) \right]
\right. \nonumber \\ & \left. 
- 3 \left( \left| X^i_j \right|^2 - \left| Y^i_j \right|^2 \right)
m_{\wt{W}_j} m_{\wt{Z}_i} \left( m^2_{\wt{W}_j} + m^2_{\wt{Z}_i} - 2 E
                                m_{\wt{W}_j} \right) \right\}.
\eeq
Here $X_j^i$ and $Y_j^i$ are the $W \wt{W}_j \wt{Z}_i$ couplings as
defined in Ref.\cite{bbkt}, and the upper integration limit $E_{\rm max}$
is given by Eq.(\ref{ea14}) with $m_{\wt{Z}_j} \rightarrow 
m_{\wt{W}_j}$ and $m_f\rightarrow 0$.

The {\em squared sneutrino exchange contribution} is given by
\be \label{ea26}
\Gamma_{\tilde \nu} = 2 \left( \wt{A}^{\nu}_{\wt{Z}_i} \right)^2
\left[ \left( \wt{A}^{\tau}_{\wt{W}_j} \right)^2 + 
     \left( B^{\tau}_{\wt{W}_j} \right)^2 \right]^2
\cdot \psi(m_{\wt{W}_j}, m_{\tilde{\nu}_{\tau}}, m_{\wt{Z}_i} ).
\ee
The couplings appearing in Eq.(\ref{ea26}) have been defined in
eqs.(\ref{ea2}), and the kinematical function $\psi$ is given in
Ref.\cite{btw2}.

The {\em pure scalar tau exchange terms} can be written as
\be \label{ea27}
\Gamma_{\tilde \tau} = \Gamma_{\tilde {\tau}_1} + \Gamma_{\tilde{\tau}_2} +
\Gamma_{\tilde{\tau}_1 \tilde{\tau}_2},
\ee
where
\ben \label{ea28} \beq
\Gamma_{\tilde{\tau}_k} &= 2 \left( \alpha_{\wt{W}_j}^{\tilde{\tau}_k} 
\right)^2 \left[ \left( \alpha_{\wt{Z}_i}^{\tilde{\tau}_k} \right)^2
+ \left( \beta_{\wt{Z}_i}^{\tilde{\tau}_k} \right)^2 \right]
\psi(m_{\wt{W}_j}, m_{\tilde{\tau}_k}, m_{\wt{Z}_i} );
\label{ea28a} \\
\Gamma_{\tilde{\tau}_1 \tilde{\tau}_2} &= 4
\alpha_{\wt{W}_j}^{\tilde{\tau}_1} \alpha_{\wt{W}_j}^{\tilde{\tau}_2} 
\left[ \alpha_{\wt{Z}_i}^{\tilde{\tau}_1} \alpha_{\wt{Z}_i}^{\tilde{\tau}_2} 
     + \beta_{\wt{Z}_i}^{\tilde{\tau}_1} \beta_{\wt{Z}_i}^{\tilde{\tau}_2} 
\right] \tilde{\psi}(m_{\wt{W}_j}, m_{\tilde{\tau}_1}, m_{\tilde{\tau}_2}, 
m_{\wt{Z}_i} ).
\label{ea28b}
\eeq \een
The couplings appearing in eqs.(\ref{ea28}) have been defined in 
eqs.(\ref{ea4})--(\ref{ea7}), and the functions $\psi, \
\tilde{\psi}$ are as defined above.

The {\em squared charged Higgs boson exchange contribution} is
\beq \label{ea29}
\Gamma_H &= \pi^2 m_{\wt{W}_j} \left( \frac {g m_{\tau} \tanb } {M_W}
\right)^2 
\int_{m_{\wt{Z}_i}}^{E_{\rm max}} dE \sqrt{ E^2 - m^2_{\wt{Z}_i} }
 \\ & \hspace*{10mm} \cdot
\frac{ \left( m^2_{\wt{W}_j} + m^2_{\wt{Z}_i} - 2 E m_{\wt{W}_j} \right)
\left\{ E \left[ \left( \alpha_{\wt{W}_j}^{(i)} \right)^2 +
 \left( \beta_{\wt{W}_j}^{(i)} \right)^2 \right] + 2 (-1)^{\theta_i+\theta_j}
m_{\wt{Z}_i}  \alpha_{\wt{W}_j}^{(i)}  \beta_{\wt{W}_j}^{(i)} \right\} }
{ \left( m^2_{\wt{W}_j} + m^2_{\wt{Z}_i} - 2 E m_{\wt{W}_j} - m^2_{H^+}
\right)^2 }. \nonumber
\eeq
Here, $E_{\rm max}$ is the same as in Eq.(\ref{ea25}), the
$H^+ \wt{W}^-_j \wt{Z}_i$ couplings have been defined in Eqs.(\ref{ea8}),
and $\theta_j \ (\equiv \theta_-$ or $\theta_+$ in the notation of
Ref.\cite{bbkt}) $= 0 \ (1)$ if the corresponding eigenvalue of the
chargino mass matrix is positive (negative).

The {\em $W-$sneutrino interference contribution} is not affected by
$\tilde{\tau}_L - \tilde{\tau}_R$ mixing and contributions $\propto f_{\tau}$;
it can be written as
\beq \label{ea30}
\Gamma_{W \tilde{\nu}} &= - 4 \sqrt{2} g^2 (-1)^{\theta_i + \theta_j}
\wt{A}^{\tau}_{\wt{W}_j} \wt{A}^{\nu}_{\wt{Z}_i} \\
&\quad
\cdot\left[\left( X^i_j - Y^i_j \right) I_1(m_{\wt{W}_j}, m_{\tilde{\nu}_{\tau}},
m_{\wt{Z}_i} ) - 
\left( X^i_j + Y^i_j \right) I_2(m_{\wt{W}_j}, m_{\tilde{\nu}_{\tau}},
m_{\wt{Z}_i} ) \right],\nonumber
\eeq
where we have introduced the functions
\ben \label{ea31} \beq
I_1(m_{\wt W}, m_{\tilde f}, m_{\wt Z} ) = \frac {\pi^2} {2 m_{\wt W}}
&\int_0^{\left( m_{\wt W} - m_{\wt Z} \right)^2} \frac {ds} {s - M_W^2}
\left[ - \frac{1}{2} Q \left( m_{\wt W} E_Q + m^2_{\tilde f} - m^2_{\wt W}
-s \right) \nonumber \right. \\ & \left. \hspace*{3mm}
- \frac{ \left( m^2_{\tilde f} - m^2_{\wt Z} \right)
         \left( m^2_{\tilde f} - m^2_{\wt W} \right) }
{4 m_{\wt W}} \log \frac { m_{\wt W} \left( E_Q + Q \right) - \mu^2 }
{ m_{\wt W} \left( E_Q - Q \right) - \mu^2 } \right] ; 
\label{ea31a} \\
I_2(m_{\wt W}, m_{\tilde f}, m_{\wt Z} ) =  \frac {\pi^2} {8 m_{\wt W}}
&\int_0^{\left( m_{\wt W} - m_{\wt Z} \right)^2} \frac {ds} {s - M_W^2}
m_{\wt Z} s \log \frac { m_{\wt W} \left( E_Q + Q \right) - \mu^2 }
{ m_{\wt W} \left( E_Q - Q \right) - \mu^2 }.
\label{ea31b}
\eeq \een
The quantities $\mu^2, \ E_Q$ and $Q$ are as in Eq.(\ref{ea19}), with
$m_{\wt{Z}_j} \rightarrow m_{\wt W}, \ m_{\wt{Z}_i} \rightarrow m_{\wt Z}$
and $m_{\tilde{f}_k} \rightarrow m_{\tilde f}$.

The same functions also appear in the {\em $W-$scalar tau interference
contributions}:
\be \label{ea32}
\Gamma_{W \tilde{\tau}} = \Gamma_{W \tilde{\tau}_1} + 
\Gamma_{W \tilde{\tau}_2},
\ee
where
\be \label{ea33}
\Gamma_{W \tilde{\tau}_k} = 4 \sqrt{2} g^2 \alpha_{\wt{W}_j}^{\tilde{\tau}_k}
\alpha_{\wt{Z}_i}^{\tilde{\tau}_k} \left[ \left( X^i_j + Y^i_j \right)
I_1( m_{\wt{W}_j}, m_{\tilde{\tau}_k}, m_{\wt{Z}_i} ) -
\left( X^i_j - Y^i_j \right)
I_2( m_{\wt{W}_j}, m_{\tilde{\tau}_k}, m_{\wt{Z}_i} ) \right].
\ee
The couplings $X^i_j, \ Y^i_j$ can be found in Ref.\cite{bbkt}; the
remaining couplings appearing in Eq.(\ref{ea33}) have been introduced in
eqs.(\ref{ea4})--(\ref{ea7}).

The {\em sneutrino--scalar tau interference terms} can be written as:
\be \label{ea34}
\Gamma_{\tilde{\nu} \tilde{\tau}} = \Gamma_{\tilde{\nu} \tilde{\tau}_1}
+ \Gamma_{\tilde{\nu} \tilde{\tau}_2},
\ee
where
\beq \label{ea35}
\Gamma_{\tilde{\nu} \tilde{\tau}_k} &= - 4 \wt{A}_{\wt{Z}_i}^{\nu}
\alpha_{\wt{W}_j}^{\tilde{\tau}_k} \\
&\quad\cdot \left[ B^{\tau}_{\wt{W}_j}
\beta_{\wt{Z}_i}^{\tilde{\tau}_k} Y(m_{\wt{W}_j}, m_{\tilde{\nu}_{\tau}},
m_{\tilde{\tau}_k}, m_{\wt{Z}_i} ) 
- (-1)^{\theta_i+\theta_j} \wt{A}^{\tau}_{\wt{W}_j}
\alpha_{\wt{Z}_i}^{\tilde{\tau}_k} \tilde{\phi} (m_{\wt{W}_j}, 
m_{\tilde{\nu}_{\tau}}, m_{\tilde{\tau}_k}, m_{\wt{Z}_i} ) \right].\nonumber
\eeq
The functions $Y, \ \tilde{\phi}$ have already been defined.

The {\em charged Higgs--sneutrino interference term} is given by:
\be \label{ea36}
\Gamma_{H \tilde{\nu}} = 2 \sqrt{2} \wt{A}^{\nu}_{\wt{Z}_i} 
B^{\tau}_{\wt{W}_j} \frac {g m_{\tau} \tanb} {m_W} 
I_H(m_{\wt{W}_j}, m_{H^+}, m_{\tilde{\nu}_{\tau}}, m_{\wt{Z}_i} ),
\ee
where we have introduced the function
\beq \label{ea37}
&I_H(m_{\wt{W}_j}, m_H, m_{\tilde f}, m_{\wt{Z}_i} ) = 
\frac {\pi^2} {2 m_{\wt{W}_j} }
\int_0^{(m_{\wt{W}_j} - m_{\wt{Z}_i})^2} \frac {ds} {s - m_H^2} 
\left\{ \frac{1}{2} s Q \beta_{\wt{W}_j}^{(i)}
\right. \\ 
& \left. \hspace*{10mm}
+ \frac{1} {4 m_{\wt{W}_j}} \left[ \beta_{\wt{W}_j}^{(i)} s m^2_{\tilde f}
+ (-1)^{\theta_i+\theta_j} \alpha_{\wt{W}_j}^{(i)} m_{\wt{W}_j} m_{\wt{Z}_i} s
\right] \log \frac { m_{\wt{W}_j} \left( E_Q + Q \right) - \mu^2 }
{ m_{\wt{W}_j} \left( E_Q - Q \right) - \mu^2 } \right\};
\nonumber
\eeq
the quantities $\mu^2, \ E_Q$ and $Q$ are as in Eq.(\ref{ea19}),
with $m_{\wt{Z}_j} \rightarrow m_{\wt{W}_j}$. The charged Higgs couplings
appearing in the integrand of Eq.(\ref{ea37}) have been defined in
eqs.(\ref{ea8}).

The same function also appears in the {\em charged Higgs--scalar tau 
interference contributions}:
\be \label{ea38}
\Gamma_{H \tilde{\tau}} = \Gamma_{H \tilde{\tau}_1} +
\Gamma_{H \tilde{\tau}_2},
\ee
where
\be \label{ea39}
\Gamma_{H \tilde{\tau}_k} = 2 \sqrt{2} \alpha_{\wt{W}_j}^{\tilde{\tau}_k}
\beta_{\wt{Z}_i}^{\tilde{\tau}_k} \frac {g m_{\tau} \tanb} {M_W}
I_H(m_{\wt{W}_j}, m_{H^+}, m_{\tilde{\tau}_k}, m_{\wt{Z}_i} ).
\ee

The partial widths for the analogous neutralino to chargino decays are
given by crossing: 
\be \label{ea40}
\Gamma(\wt{Z}_i \rightarrow \wt{W}_j^+ \tau^- \bar{\nu}_{\tau})
= \Gamma(\wt{W}_j^- \rightarrow \wt{Z}_i \tau^- \bar{\nu}_{\tau})
( m_{\wt{W}_j} \leftrightarrow m_{\wt{Z}_i} ).
\ee
Note that $\wt{Z}_i$ can also decay into $\wt{W}_j^- \tau^+ \nu_{\tau}$
final states, with equal probability. However, these neutralino 
decays are usually not very important, since they are either phase space 
suppressed, or have to compete with 2--body decays of the heavy neutralinos.

\subsection{$\tilde g \rightarrow \wt{W}_i t \bar{b}$ Decays}

These decays proceed through the exchange of any of the four stop and
sbottom mass eigenstates; in the limit $\theta_b, \ f_b \rightarrow 0$
considered in the existing literature \cite{btw2}, only one of the
two sbottom eigenstates contributes here, since $\tilde{b}_R$ does not
couple to charginos in this limit. Fortunately the general case does not
introduce terms with new Dirac structure in the matrix elements; the
necessary phase space integrals can therefore be extracted from the
Appendix of Ref.\cite{btw2}.

We begin by defining eight kinematical functions:
\ben \label{ea41} \beq
G_1(m_{\tilde g}, m_{\tilde t}, m_{\wt W}) &= m_{\tilde g}
\int \frac {d E_t p_t E_t \left( m^2_{\tilde g} + m^2_t
- 2 E_t m_{\tilde g} - m^2_{\wt W} \right)^2 }
{ \left( m^2_{\tilde g} + m^2_t - 2 E_t m_{\tilde g} - m^2_{\tilde t} 
\right)^2 \left(  m^2_{\tilde g} + m^2_t - 2 E_t m_{\tilde g} \right)};
\label{ea41a} \\
G_2(m_{\tilde g}, m_{\tilde b}, m_{\wt W}) &=
m_{\tilde g} \int dE_{\bar b} E^2_{\bar b} \lambda^{1/2}(m^2_{\tilde g}
+m_b^2-2 E_{\bar b} m_{\tilde g}, m^2_{\wt W}, m^2_t)
\nonumber \\ & \hspace*{15mm} \cdot
\frac {  m^2_{\tilde g}+m_b^2 - m^2_t - 2 E_{\bar b} m_{\tilde g} - m^2_{\wt W} }
{ \left( m^2_{\tilde g}+m_b^2 - 2 E_{\bar b} m_{\tilde g} - m^2_{\tilde b} 
\right)^2 \left(  m^2_{\tilde g}+m_b^2 - 2 E_{\bar b} m_{\tilde g} \right)};
\label{ea41b} \\
G_3(m_{\tilde g}, m_{\tilde b}, m_{\wt W}) &=
(-1)^{\theta_{\wt W}} 4 m_{\tilde g} m_{\wt W} m_t 
\int dE_{\bar b} E^2_{\bar b} \lambda^{1/2}(m^2_{\tilde g}+m_b^2
-2 E_{\bar b} m_{\tilde g}, m^2_{\wt W}, m^2_t)
\nonumber \\ & \hspace*{15mm} \cdot
\frac {  1 }
{ \left( m^2_{\tilde g}+m_b^2 - 2 E_{\bar b} m_{\tilde g} - m^2_{\tilde b} 
\right)^2 \left(  m^2_{\tilde g}+m_b^2 - 2 E_{\bar b} m_{\tilde g} \right)};
\label{ea41c} \\
G_4(m_{\tilde g}, m_{\tilde t}, m_{\tilde b}, m_{\wt W}) &= 
(-1)^{\theta_{\tilde g} + \theta_{\wt W} } m_{\tilde g} m_{\wt W}
\int \frac {d E_t }
{ m^2_{\tilde g} + m^2_t - 2 E_t m_{\tilde g} - m^2_{\tilde t} } 
\\ & \hspace*{15mm} 
\cdot \left[ E_{\bar b}({\rm max}) - E_{\bar b}({\rm min}) -
\frac{ m^2_{\tilde b} + m^2_t - 2 E_t m_{\tilde g} - m^2_{\wt W} } 
{2 m_{\tilde g}} \log X \right];
\label{ea41d} \nonumber \\
G_5(m_{\tilde g}, m_{\tilde t}, m_{\tilde b}, m_{\wt W}) &= 
(-1)^{\theta_{\tilde g}} \frac {m_t}{2} 
\int d E_t 
\frac {m^2_{\tilde g} + m^2_t - 2 E_t m_{\tilde g} - m^2_{\wt W} } 
{ m^2_{\tilde g} + m^2_t - 2 E_t m_{\tilde g} - m^2_{\tilde t} } \log X ;
\label{ea41e} \\
G_6(m_{\tilde g}, m_{\tilde t}, m_{\tilde b}, m_{\wt W}) &= \frac{1}{2}
\int \frac {d E_t }
{ m^2_{\tilde g} + m^2_t - 2 E_t m_{\tilde g} - m^2_{\tilde t} } 
\cdot \left\{\left[ m_{\tilde g} \left( m^2_{\tilde g} + m^2_t - 
2 E_t m_{\tilde g} - m^2_{\wt W} \right)
\nonumber \right. \right. \\ & \left. \left. \hspace*{35mm} 
- \frac {m^2_{\tilde b} - m^2_{\tilde g} } {m_{\tilde g}} \left(
2 E_t m_{\tilde g} -m^2_t - m^2_{\tilde g} \right) \right] \log X
\right. \nonumber \\ & \left. \hspace*{25mm}
+ 2 \left( 2 E_t m_{\tilde g} - m^2_t - m^2_{\tilde g} \right)
\left[  E_{\bar b}({\rm max}) - E_{\bar b}({\rm min}) \right] \right\};
\label{ea41f} \\
G_7(m_{\tilde g}, m_{\tilde t}, m_{\tilde b}, m_{\wt W}) &= 
(-1)^{\theta_{\wt W}}\frac{1}{2} m_{\wt W} m_t
\int \frac {d E_t }
{ m^2_{\tilde g} + m^2_t - 2 E_t m_{\tilde g} - m^2_{\tilde t} } 
\nonumber \\ & \hspace*{15mm} 
\cdot \left\{ 2 \left[ E_{\bar b}({\rm max}) - E_{\bar b}({\rm min}) \right]
- \frac{ m^2_{\tilde b} - m^2_{\tilde g} } {m_{\tilde g}} \log X \right\};
\label{ea41g} \\
G_8(m_{\tilde g}, m_{\tilde{t}_1}, m_{\tilde{t}_2}, m_{\wt W}) &= 
(-1)^{\theta_{\tilde g}} m_t m_{\tilde g} \nonumber \\
& \cdot \int d E_t 
\frac { \left( m^2_{\tilde g} + m^2_t - 2 E_t m_{\tilde g} - m^2_{\wt W} 
\right) \left[ E_{\bar b}({\rm max}) - E_{\bar b}({\rm min}) \right]} 
{ \left(m^2_{\tilde g} + m^2_t - 2 E_t m_{\tilde g} - m^2_{\tilde{t}_1} \right)
 \left(m^2_{\tilde g} + m^2_t - 2 E_t m_{\tilde g} - m^2_{\tilde{t}_2} \right)
 }.
\label{ea41h}
\eeq \een
Here, $\theta_{\tilde g} = 0$ (1) if a positive (negative) gluino mass
parameter is chosen, and $\theta_{\wt W}$ ($=\theta_-$ or $\theta_+$ in the
notation of Ref.\cite{bbkt}) is the corresponding quantity
for the chargino mass eigenstate. Further, we have introduced
$E_{\bar b}({\rm min, \ max})$\cite{btw2}, which are given by
\beq
\frac  
{(m^2_{\tilde g} + m^2_t - 2 m_{\tilde g}E_t + m_b^2 - m^2_{\wt W})
(m_{\tilde g}-E_t) \mp p_t\lambda^{1/2}(m_{\tilde g}^2+m_t^2-2m_{\tilde g} E_t,
m_b^2,m_{\wt W}^2) }
{2 \left( m_{\tilde g}^2+m_t^2 -2 E_t m_{\tilde g} \right) }.
\eeq
Also,
\ben \label{ea42} \beq
p_t =& \sqrt{E_t^2 - m_t^2 } \hspace{1cm} {\rm and} \label{ea42b} \\
X =& \frac { m^2_{\tilde b} + 2 E_{\bar b}({\rm max}) m_{\tilde g}
- m^2_{\tilde g} }
{ m^2_{\tilde b} + 2 E_{\bar b}({\rm min}) m_{\tilde g} - m^2_{\tilde g} }.
\label{ea42c}
\eeq \een
Finally, the limits of integration over $E_t$ in eqs.(\ref{ea41}) are from 
$m_t$ to
$\left( m^2_{\tilde g} + m_t^2 - (m_{\wt W}+m_b)^2 \right) / 2 m_{\tilde g}$,
while the integration over $E_{\bar b}$ in eqs.(\ref{ea41}b,c) goes
from $m_b$ to $\left[ m^2_{\tilde g}-\left( m_t+m_{\wt W} \right)^2 \right]
/ 2 m_{\tilde g}$.

The partial widths for the processes under consideration can be written as
\be \label{ea43}
\Gamma(\tilde{g} \rightarrow t \bar{b} \wt{W}_i) = \frac {1} 
{\left( 2 \pi \right)^2} \frac {1} {2 m_{\tilde g}} \pi^2 g_s^2 \left(
\Gamma_{\tilde{t}_1} + \Gamma_{\tilde{t}_2} + \Gamma_{\tilde{t}_1 \tilde{t}_2}
 + \Gamma_{\tilde{b}_1} + \Gamma_{\tilde{b}_2} 
+ \sum_{k,l=1}^2 \Gamma_{\tilde{t}_k \tilde{b}_l} \right),
\ee
where $g_s$ is the $SU(3)_c$ gauge coupling. Note that in the limit
$m_b \rightarrow 0$ the two sbottom exchange diagrams do not interfere
with each other. The individual contributions in Eq.(\ref{ea43}) are
given by:
\ben \label{ea44} \beq
\Gamma_{\tilde{t}_k} &= \left[ \left( \alpha_{\wt{W}_i}^{\tilde{t}_k}
\right)^2 + \left( \beta_{\wt{W}_i}^{\tilde{t}_k} \right)^2 \right]
\left[ G_1(m_{\tilde g}, m_{\tilde{t}_k}, m_{\wt{W}_i}) 
\nonumber \right. \\ & \left. \hspace*{41mm}
- (-1)^k \sin(
2 \theta_t) G_8(m_{\tilde g}, m_{\tilde{t}_k}, m_{\tilde{t}_k}, m_{\wt{W}_i})
\right];
\label{ea44a} \\
\Gamma_{\tilde{t}_1 \tilde{t}_2} &= -2 \left( \alpha_{\wt{W}_i}^{\tilde{t}_1}
 \alpha_{\wt{W}_i}^{\tilde{t}_2} + \beta_{\wt{W}_i}^{\tilde{t}_1}
\beta_{\wt{W}_i}^{\tilde{t}_2 } \right) \cos(2 \theta_t)
G_8(m_{\tilde g}, m_{\tilde{t}_1}, m_{\tilde{t}_2}, m_{\wt{W}_i});
\label{ea44b} \\
\Gamma_{\tilde{b}_k} &= \left[ \left( \alpha_{\wt{W}_i}^{\tilde{b}_k}
\right)^2 + \left( \beta_{\wt{W}_i}^{\tilde{b}_k} \right)^2 \right]
G_2(m_{\tilde g}, m_{\tilde{b}_k}, m_{\wt{W}_i}) -
 \alpha_{\wt{W}_i}^{\tilde{b}_k} \beta_{\wt{W}_i}^{\tilde{b}_k}
G_3(m_{\tilde g}, m_{\tilde{b}_k}, m_{\wt{W}_i}) ;
\label{ea44c} \\
\Gamma_{\tilde{t}_1 \tilde{b}_1} &= \left( \cos \! \theta_t \sin \! \theta_b
\alpha_{\wt{W}_i}^{\tilde{b}_1} \beta_{\wt{W}_i}^{\tilde{t}_1} 
+ \sin \! \theta_t \cos \! \theta_b \beta_{\wt{W}_i}^{\tilde{b}_1}
 \alpha_{\wt{W}_i}^{\tilde{t}_1} \right)
G_6(m_{\tilde g}, m_{\tilde{t}_1}, m_{\tilde{b}_1}, m_{\wt{W}_i})
\nonumber \\ &
- \left( \cos \! \theta_t \cos \! \theta_b
\alpha_{\wt{W}_i}^{\tilde{b}_1} \alpha_{\wt{W}_i}^{\tilde{t}_1} 
+ \sin \! \theta_t \sin \! \theta_b \beta_{\wt{W}_i}^{\tilde{b}_1}
 \beta_{\wt{W}_i}^{\tilde{t}_1} \right)
G_4(m_{\tilde g}, m_{\tilde{t}_1}, m_{\tilde{b}_1}, m_{\wt{W}_i})
\nonumber \\ &
+ \left( \cos \! \theta_t \cos \! \theta_b
\beta_{\wt{W}_i}^{\tilde{b}_1} \alpha_{\wt{W}_i}^{\tilde{t}_1} 
+ \sin \! \theta_t \sin \! \theta_b \alpha_{\wt{W}_i}^{\tilde{b}_1}
 \beta_{\wt{W}_i}^{\tilde{t}_1} \right)
G_5(m_{\tilde g}, m_{\tilde{t}_1}, m_{\tilde{b}_1}, m_{\wt{W}_i})
\nonumber \\ &
- \left( \cos \! \theta_t \sin \! \theta_b
\beta_{\wt{W}_i}^{\tilde{b}_1} \beta_{\wt{W}_i}^{\tilde{t}_1} 
+ \sin \! \theta_t \cos \! \theta_b \alpha_{\wt{W}_i}^{\tilde{b}_1}
 \alpha_{\wt{W}_i}^{\tilde{t}_1} \right)
G_7(m_{\tilde g}, m_{\tilde{t}_1}, m_{\tilde{b}_1}, m_{\wt{W}_i}).
\label{ea44d}
\eeq \een
The couplings appearing in eqs.(\ref{ea44}) are listed in
eqs.(\ref{ea4})--(\ref{ea7}). 
The other stop--sbottom interference contributions can be obtained from
Eq.(\ref{ea44d}) by substituting the appropriate coupling constants and
squark masses; in addition, one has to apply the substitution rules
(\ref{ea6}) to the factors in Eq.(\ref{ea44d}) that depend on third
generation squark mixing angles. Finally, we note that gluinos have the same
partial widths for decays into $t \bar{b} \wt{W}_i^-$ and
$\bar{t} b \wt{W}_i^+$ final states.

These formulae have been incorporated into the event generator
ISAJET 7.32\cite{isajet}. Finally, we remark that we have also updated the
formula for $\Gamma(\tg \to t\bar{t} \tz_i)$ that appears in Ref.\cite{btw2}
to include $\tt_L-\tt_R$ mixing effects. This has also been
incorporated into ISAJET.
%

\newpage
%
%

\iftightenlines\else\newpage\fi
\iftightenlines\global\firstfigfalse\fi
\def\dofig#1#2{\epsfxsize=#1\centerline{\epsfbox{#2}}}



\begin{figure}
\caption[]{Selected sparticle and Higgs boson masses
versus $\tan\beta$ for the mSUGRA model for
parameters {\it a}) ($m_0,m_{1/2},A_0)=(150,150,0)$ GeV and
{\it b}) ($m_0,m_{1/2},A_0)=(150,500,0)$ GeV, for both signs of the
parameter $\mu$. We take $m_t=170$ GeV.}
\label{nfig1}
\end{figure}

\begin{figure}
\caption[]{Chargino ($\tw_1$) and 
neutralino ($\tz_2$) branching fractions versus $\tan\beta$. In
{\it a}) and {\it b}), we take the parameters
($m_0,m_{1/2},A_0)=(150,150,0)$ GeV while in {\it c}) and {\it d}) we take
($m_0,m_{1/2},A_0)=(150,500,0)$ GeV. In all frames, $\mu >0$ and
$m_t=170$ GeV.}
\label{nfig2}
\end{figure}

\begin{figure}
\caption[]{Gluino and squark mass contours in the
$m_0\ vs.\ m_{1/2}$ parameter plane, for {\it a}) $\tan\beta =2$,
{\it b}) $\tan\beta =20$, {\it c}) $\tan\beta =35$ and 
{\it d}) $\tan\beta =45$. In all frames, we take $A_0=0$, $\mu >0$
and $m_t=170$ GeV. The bricked regions are excluded by theoretical constraints,
while the gray regions are excluded by LEP2 bounds on $m_{\tw_1}$.}
\label{nfig3}
\end{figure}

\begin{figure}
\caption[]{A plot of points accessible to Tevatron MI and TeV33
searches for mSUGRA via $\eslt +$ multijet events. A $5\sigma$
signal above background is found for some value of $E_T^c$ for the MI
for gray squares, while white squares are accessible only at TeV33.
Points with a $\times$ symbol are inaccessible to MI and TeV33.}
\label{nfig4}
\end{figure}

\begin{figure}
\caption[]{Same as Fig.~\protect\ref{nfig4}, except we require in addition
that at least one of the jets be an identified $b$-jet.}
\label{nfig5}
\end{figure}

\begin{figure}
\caption[]{A plot of the reach of the Tevatron MI and TeV33 for mSUGRA via
the JOS signal.}
\label{nfig6}
\end{figure}

\begin{figure}
\caption[]{A plot of the reach of the Tevatron MI and TeV33 for mSUGRA via
the J3L signal.}
\label{nfig7}
\end{figure}

\begin{figure}
\caption[]{A plot of the reach of the Tevatron MI and TeV33 for mSUGRA via
the C3L signal.}
\label{nfig8}
\end{figure}

\begin{figure}
\caption[]{A plot of the reach of the Tevatron MI and TeV33 for mSUGRA via
the C3LT signal.}
\label{nfig9}
\end{figure}


\begin{figure}
\caption[]{A plot of the combined reach of the Tevatron MI and TeV33 for 
mSUGRA via {\it all} of the signal channels considered in this paper.}
\label{nfig11}
\end{figure}
\vfill\eject

\centerline{\epsfbox{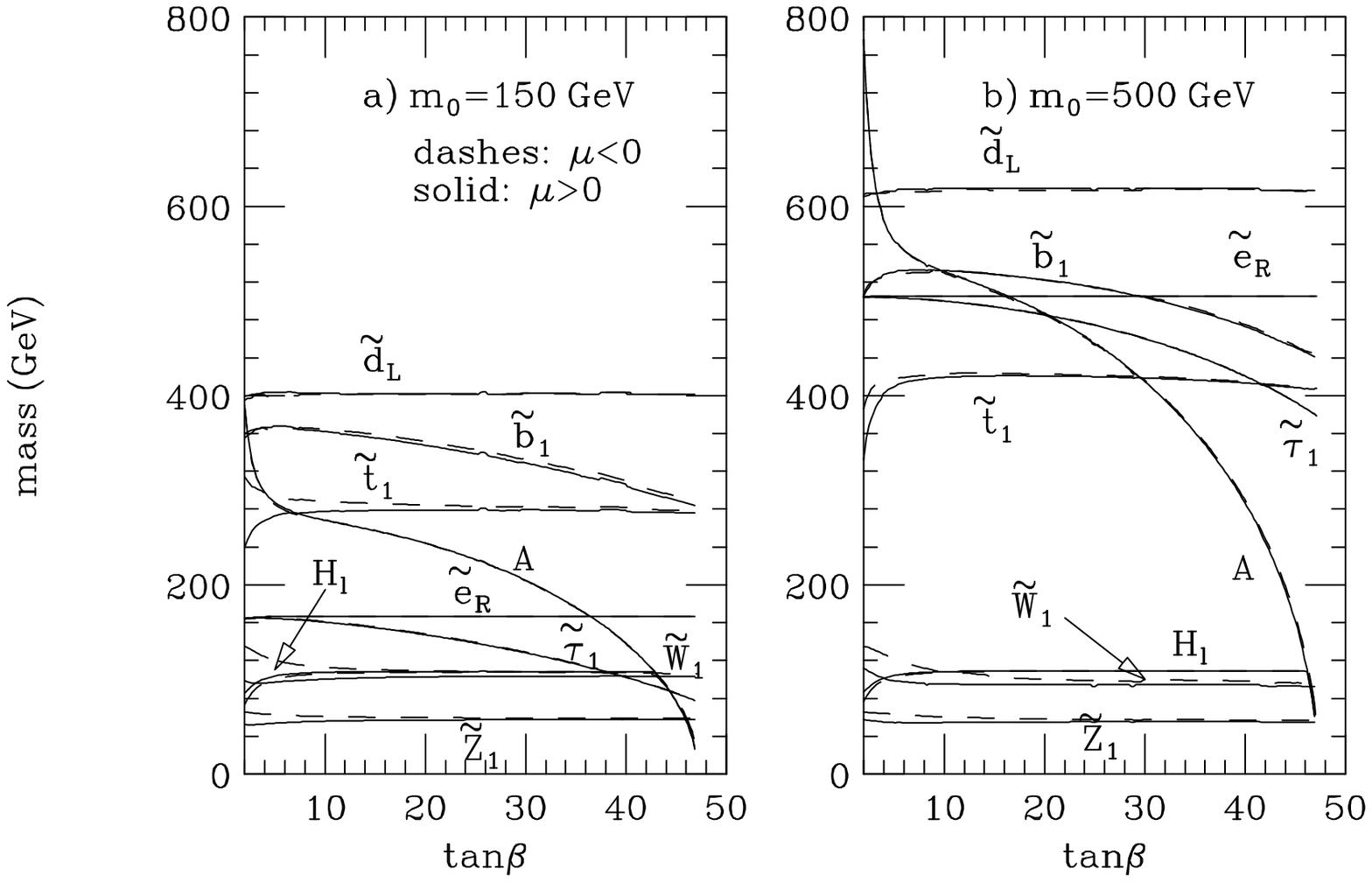}}
\smallskip
\centerline{Fig.~1}
\vfill\eject

\centerline{\epsfbox{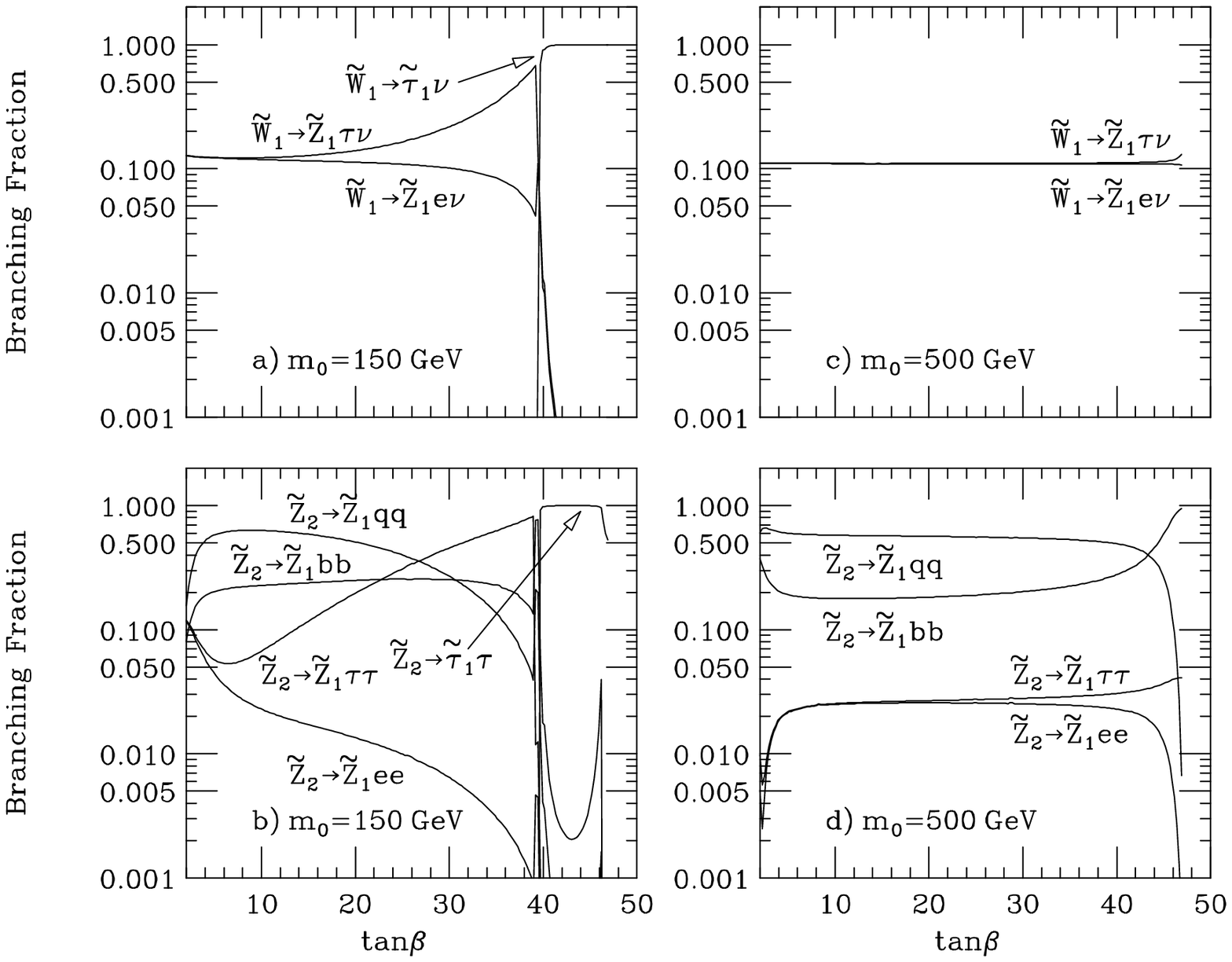}}
\smallskip
\centerline{Fig.~2}
\vfill\eject

\centerline{\epsfbox{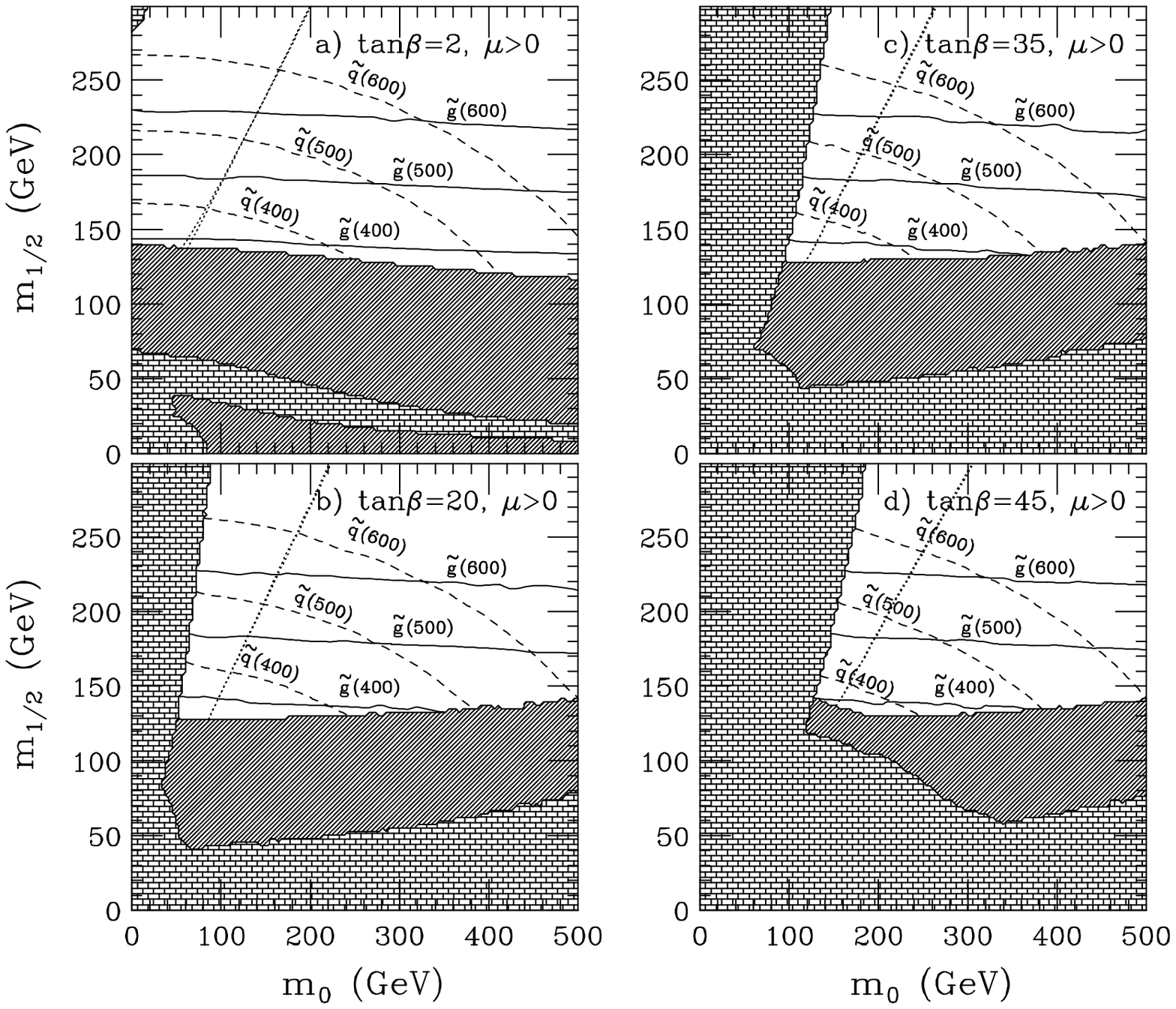}}
\smallskip
\centerline{Fig.~3}
\vfill\eject

\centerline{\epsfbox{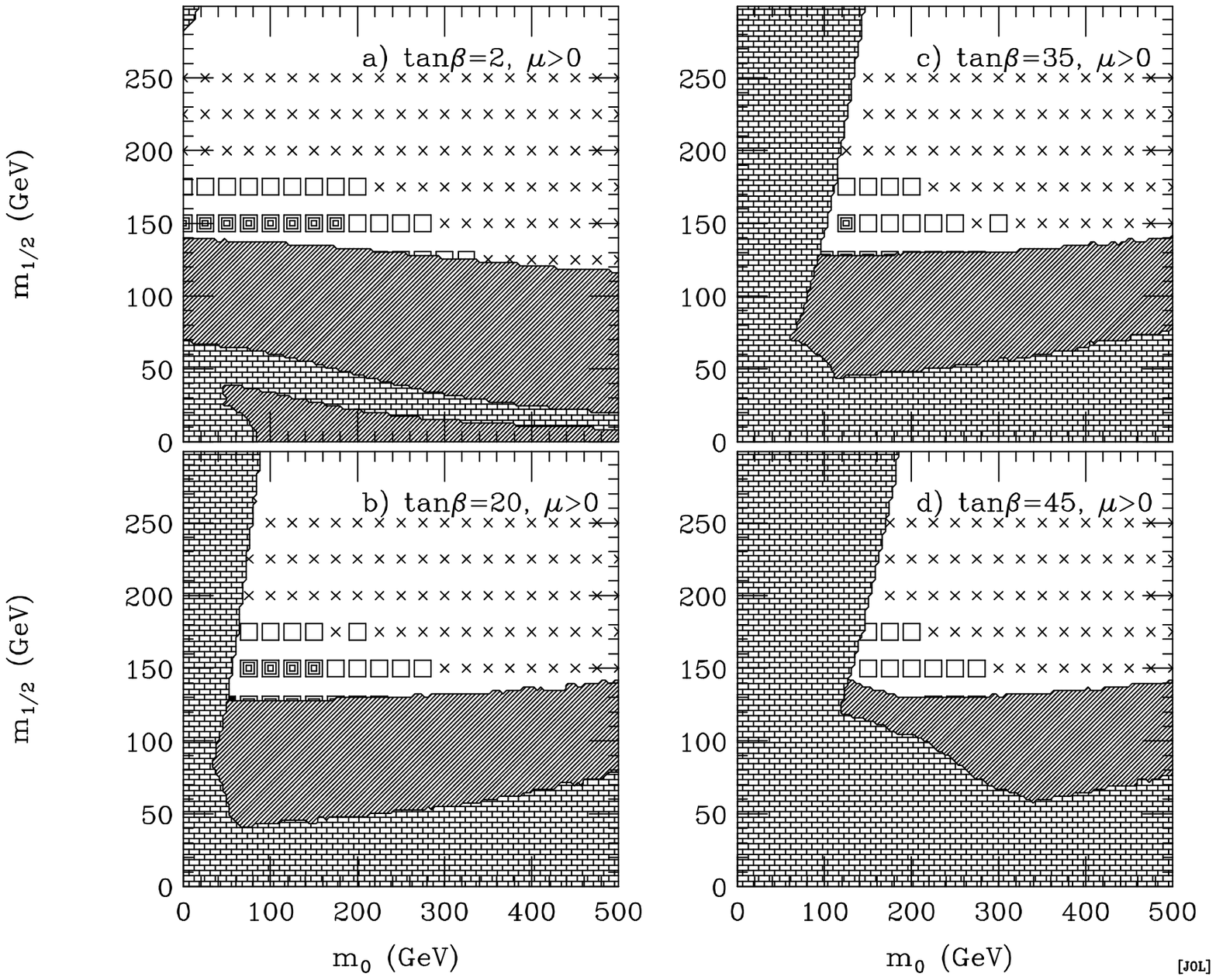}}
\smallskip
\centerline{Fig.~4}
\vfill\eject

\centerline{\epsfbox{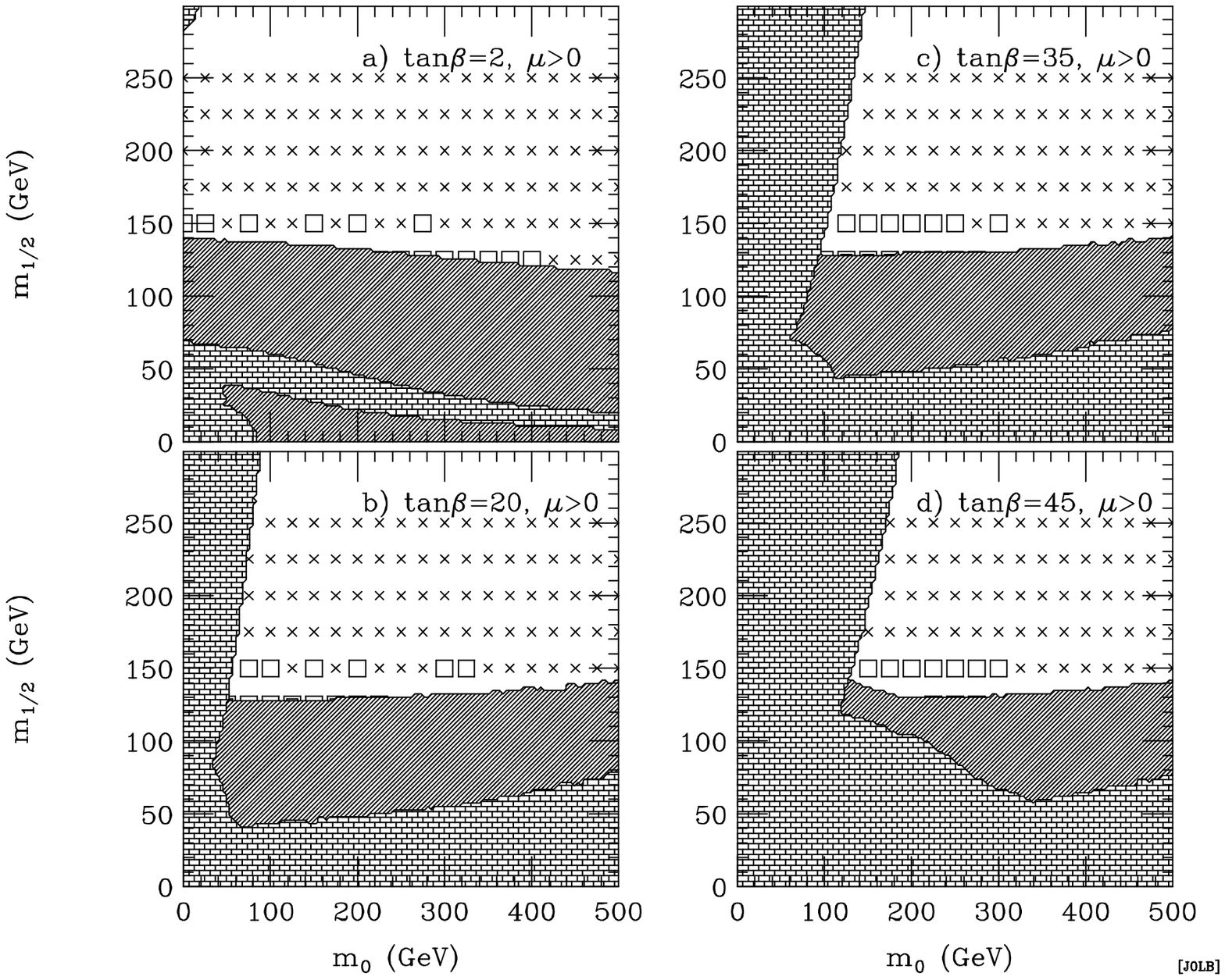}}
\smallskip
\centerline{Fig.~5}
\vfill\eject

\centerline{\epsfbox{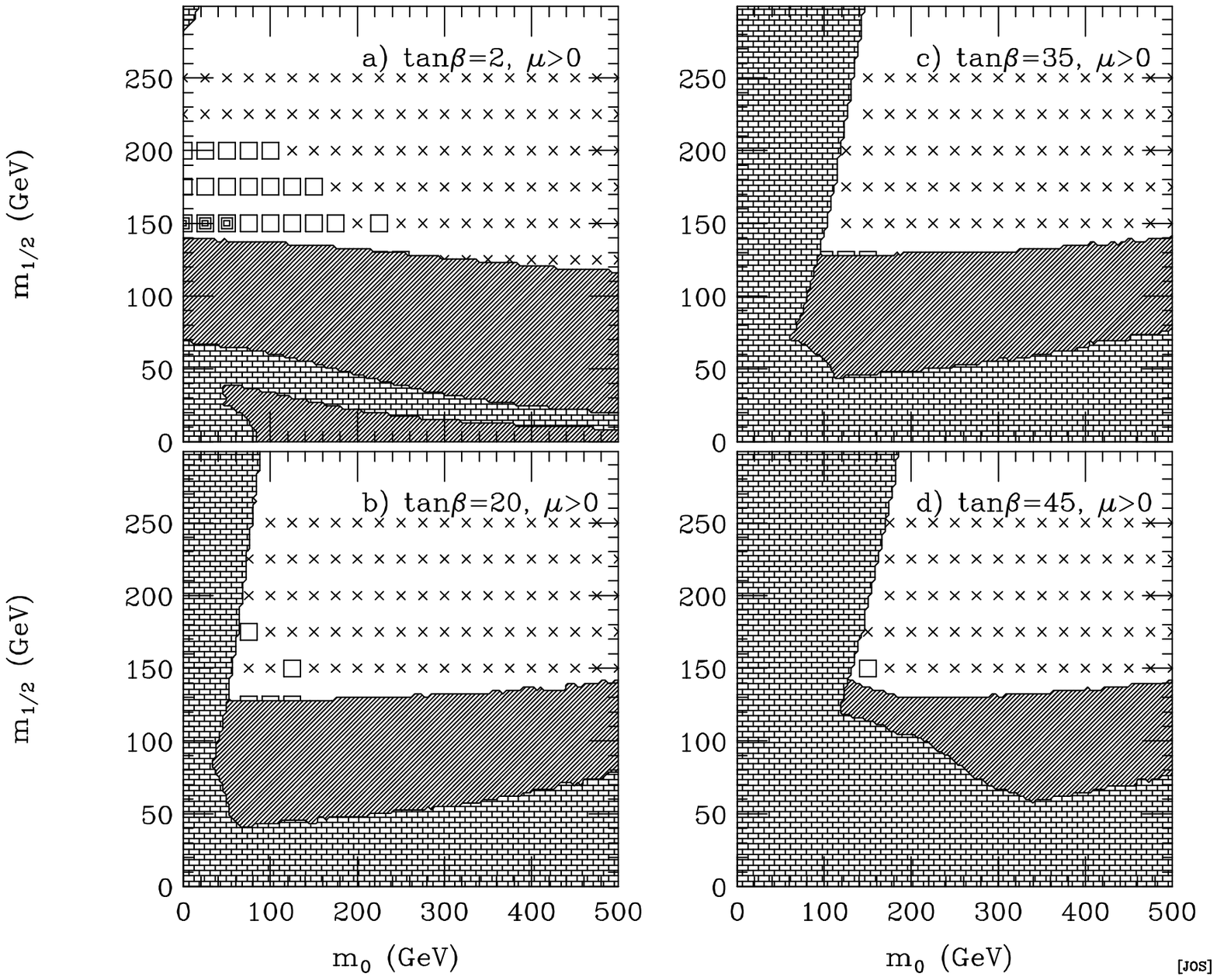}}
\smallskip
\centerline{Fig.~6}
\vfill\eject

\centerline{\epsfbox{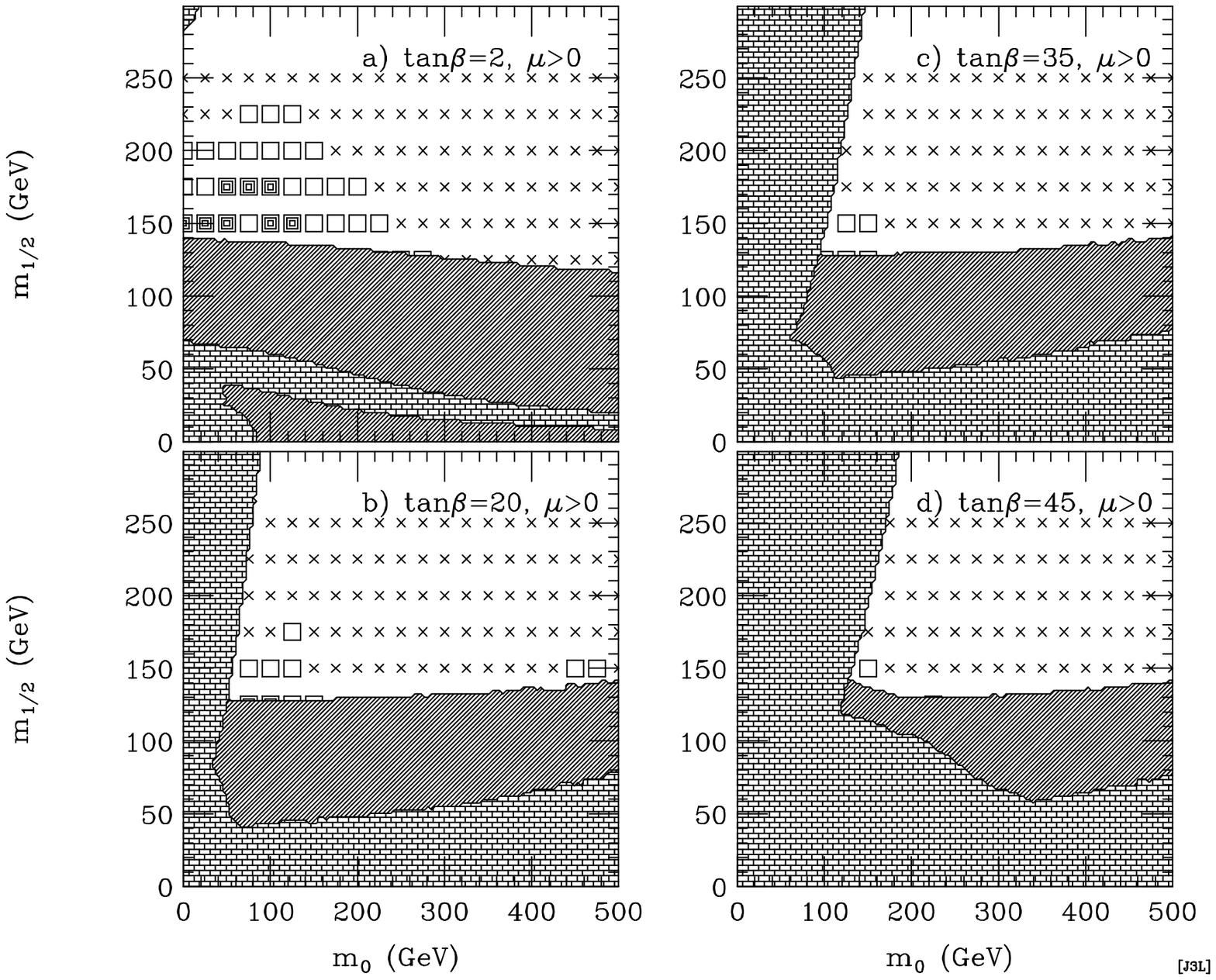}}
\smallskip
\centerline{Fig.~7}
\vfill\eject

\centerline{\epsfbox{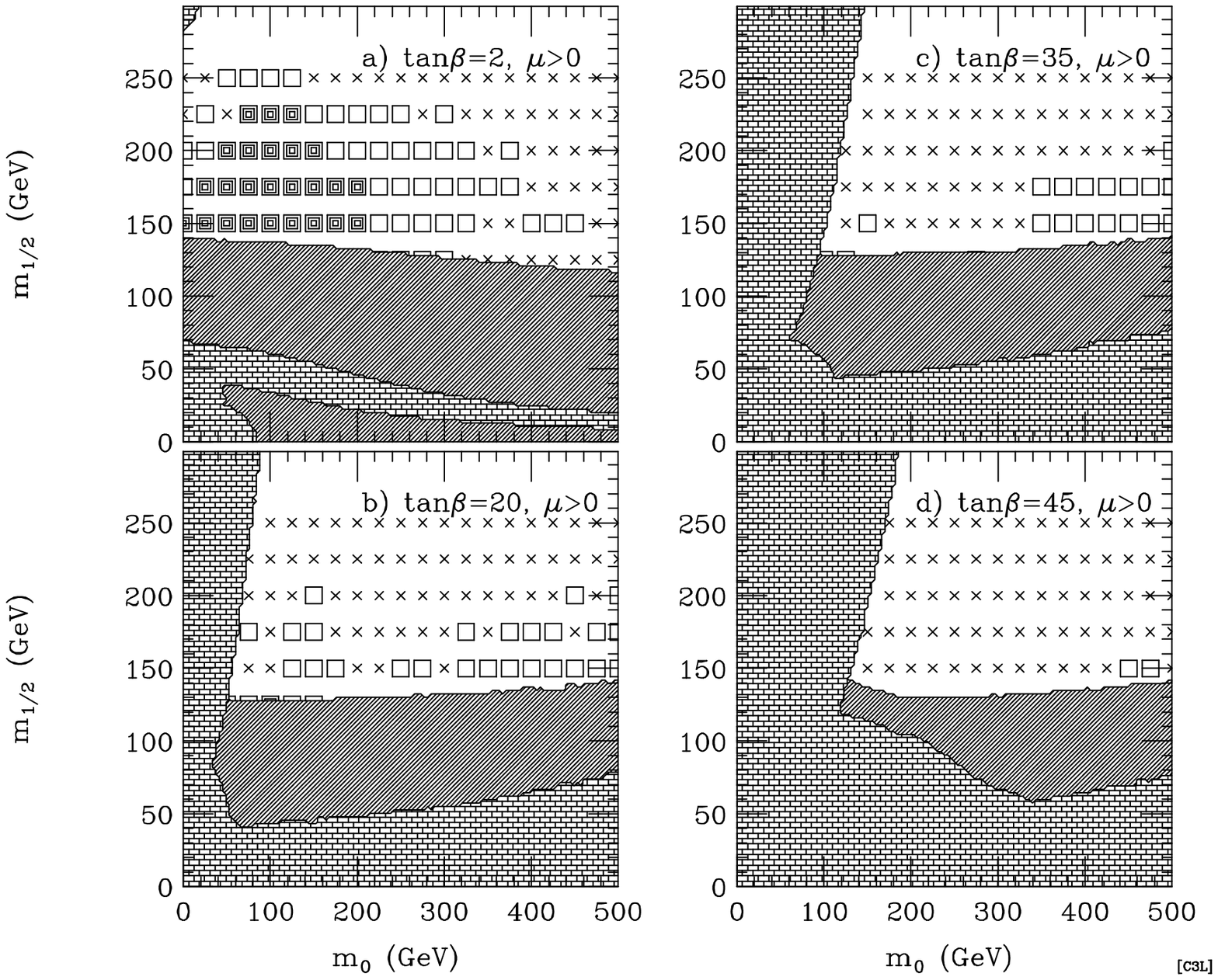}}
\smallskip
\centerline{Fig.~8}
\vfill\eject

\centerline{\epsfbox{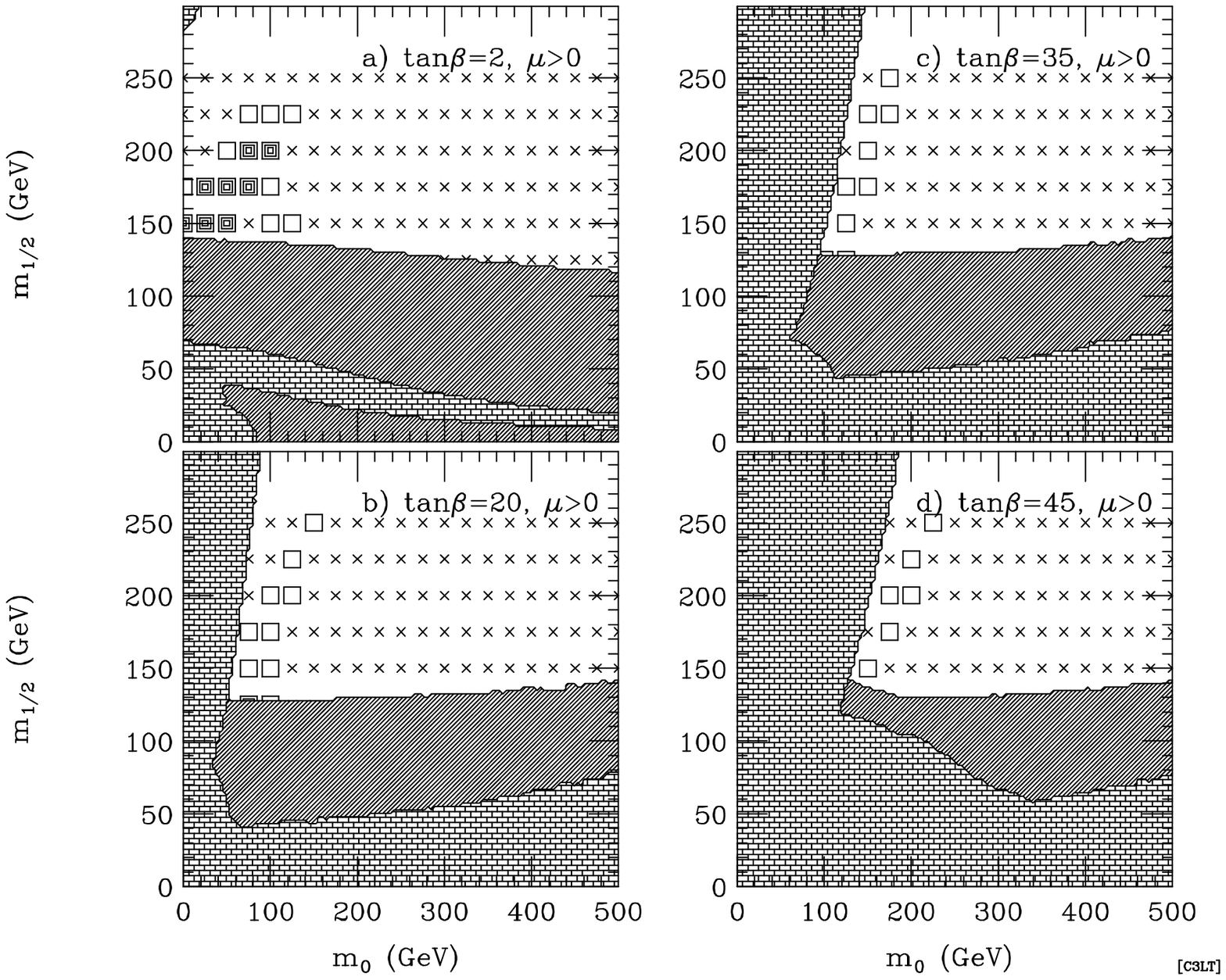}}
\smallskip
\centerline{Fig.~9}
\vfill\eject

\centerline{\epsfbox{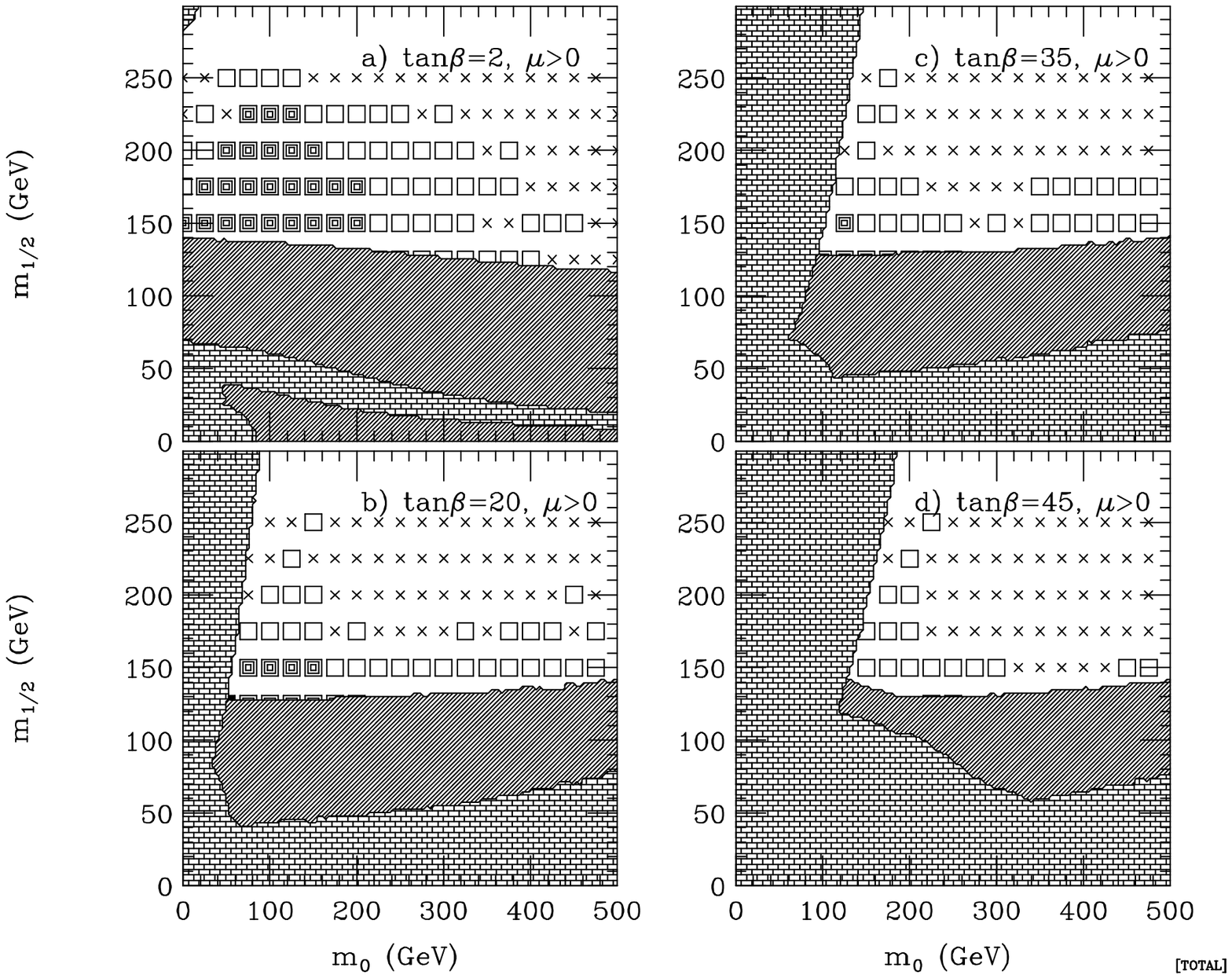}}
\smallskip
\centerline{Fig.~10}
\vfill\eject

\end{document}